\documentclass[sn-mathphys-num]{sn-jnl}

\usepackage{graphicx}%
\usepackage{multirow}%
\usepackage{amsmath}
\usepackage{amssymb,amsfonts,amstext}%
\usepackage[title]{appendix}%
\usepackage{xcolor}%
\usepackage{textcomp}%
\usepackage{manyfoot}%
\usepackage{booktabs}%
\usepackage{algorithm}%
\usepackage{algorithmicx}%
\usepackage{algpseudocode}%
\usepackage{listings}%

\usepackage{lineno,hyperref}
\usepackage{subcaption}
\usepackage{forest}
\usepackage{courier}
\usepackage{bbold}
\usetikzlibrary{shapes,arrows,positioning,automata}

\tikzset{
  treenode/.style = {shape=rectangle, rounded corners,
                     draw, align=center,
                     top color=white,
                     bottom color=white},
  root/.style     = {treenode, font=\Large,
                     bottom color=red!30},
  env/.style      = {treenode, font=\ttfamily\normalsize},
  dummy/.style    = {circle,draw}
}
\usepackage{caption}
\usepackage{listings}
\usepackage{xcolor} 
\usepackage{threeparttable}
\usepackage[usestackEOL]{stackengine}
\usepackage{xcolor}
\usepackage[english]{babel}
\usepackage[utf8]{inputenc}
\edef\tmp{\the\baselineskip}
\setstackgap{L}{\tmp}
\modulolinenumbers[1]

\DeclareMathOperator*{\argmax}{arg\,max}
\DeclareMathOperator*{\argmin}{arg\,min}

\begin{document}

\title[MOOSE ProbML]{MOOSE ProbML: Parallelized Probabilistic Machine Learning and Uncertainty Quantification for Computational Energy Applications}


\author*[1,6]{\fnm{Somayajulu L. N.} \sur{Dhulipala}}\email{Som.Dhulipala@inl.gov}

\author[2]{\fnm{Peter} \sur{German}}

\author[3,1]{\fnm{Yifeng} \sur{Che}}

\author[2]{\fnm{Zachary M.} \sur{Prince}}

\author[4]{\fnm{Xianjian} \sur{Xie}}

\author[1]{\fnm{Pierre-Cl\'ement A.} \sur{Simon}}

\author[5]{\fnm{Vincent M.} \sur{Labour\'e}}

\author[4]{\fnm{Hao} \sur{Yan}}

\affil*[1]{\orgdiv{Computational Mechanics and Materials Department}, \orgname{Idaho National Laboratory}, \orgaddress{\city{Idaho Falls}, \postcode{83415}, \state{ID}, \country{USA}}}

\affil[2]{\orgdiv{Computational Frameworks Department}, \orgname{Idaho National Laboratory}, \orgaddress{\city{Idaho Falls}, \postcode{83415}, \state{ID}, \country{USA}}}

\affil[3]{\orgdiv{Woodruff School of Mechanical Engineering}, \orgname{Georgia Institute of Technology}, \orgaddress{ \city{Atlanta}, \postcode{30332}, \state{GA}, \country{USA}}}

\affil[4]{\orgdiv{School of Computing and Augmented Intelligence}, \orgname{Arizona State University}, \orgaddress{ \city{Tempe}, \postcode{85287}, \state{AZ}, \country{USA}}}

\affil[5]{\orgdiv{Reactor Physics Methods and Analysis Department}, \orgname{Idaho National Laboratory}, \orgaddress{\city{Idaho Falls}, \postcode{83415}, \state{ID}, \country{USA}}}

\affil[6]{\orgdiv{Civil and Environmental Engineering Department}, \orgname{Idaho State University}, \orgaddress{\city{Pocatello}, \postcode{83209}, \state{ID}, \country{USA}}}


\abstract{This paper presents the development and demonstration of massively parallel probabilistic machine learning (ML) and uncertainty quantification (UQ) capabilities within the Multiphysics Object-Oriented Simulation Environment (MOOSE), an open-source computational platform for parallel finite element and finite volume analyses. In addressing the computational expense and uncertainties inherent in complex multiphysics simulations, this paper integrates Gaussian process (GP) variants, active learning, Bayesian inverse UQ, adaptive forward UQ, Bayesian optimization, evolutionary optimization, and Markov chain Monte Carlo (MCMC) within MOOSE. It also elaborates on the interaction among key MOOSE systems---\texttt{Sampler}, \texttt{MultiApp}, \texttt{Reporter}, and \texttt{Surrogate}---in enabling these capabilities. The modularity offered by these systems enables development of a multitude of probabilistic ML and UQ algorithms in MOOSE. Example code demonstrations include parallel active learning and parallel Bayesian inference via active learning. The impact of these developments is illustrated through five applications relevant to computational energy applications: UQ of nuclear fuel fission product release, using parallel active learning Bayesian inference; very rare events analysis in nuclear microreactors using active learning; advanced manufacturing process modeling using multi-output GPs (MOGPs) and dimensionality reduction; fluid flow using deep GPs (DGPs); and tritium transport model parameter optimization for fusion energy, using batch Bayesian optimization.}


\maketitle


\section{Introduction}


The Multiphysics Object-Oriented Simulation Environment (MOOSE), an open-source computational platform for parallel finite element and finite volume analyses, is being developed and maintained primarily at Idaho National Laboratory, and has a wide user and developer base spanning academia, industry, and national laboratories \cite{Giudicelli2024a}. It is easy to install, offers extensive tutorials, comes with built-in physics modules, and naturally lends itself to multiscale and multiphysics simulations. MOOSE supports a vibrant community of computational scientists and engineers via a highly active discussions forum, and its code base receives tens of pull requests each month (\href{https://github.com/idaholab/moose}{https://github.com/idaholab/moose}). MOOSE has traditionally supported computational simulations intended to advance energy solutions such as nuclear fission energy, geothermal energy, and, more recently, nuclear fusion energy. Several applications were built by using MOOSE to tackle specific problems such as nuclear fuel performance (BISON \cite{WilliamsonBISON}), structural materials aging (Grizzly \cite{Spencer2021x}), medium-fidelity thermal hydraulics (Pronghorn \cite{Novak2021x}), radiation transport (Griffin \cite{Wang2025x}), seismic analysis (Mastodon \cite{Veeraraghavan2021x}), mesoscale materials simulations (Marmot \cite{Tonks2012x}), high-fidelity thermal-hydraulics and/or radiation transport (Cardinal \cite{Novak2022x}), tritium transport for fusion energy (TMAP8 \cite{Simon2025}), thermal-hydraulic-mechanical-chemical processes in geothermal systems (Falcon \cite{Podgorney2021x}), etc. MOOSE also provides a stochastic tools module to support uncertainty quantification (UQ) and propagation, as well as surrogate model development for multiphysics simulations \cite{Slaughter2023a}. This paper presents the development and demonstration of massively parallel probabilistic machine learning (ML) and UQ in the stochastic tools module to support capabilities such as Gaussian process (GP) ML, active learning, Bayesian inference, rare events analysis, Bayesian optimization, and evolutionary optimization. These capabilities in MOOSE are motivated by the following: (1) complex multiphysics simulations, when validated with experimental data, are subject to different sources of uncertainties (i.e., model parameters, model inadequacy, and experimental noise) that must be quantified and propagated to the outputs; (2) complex multiphysics models are computationally expensive to run, especially in a UQ setting, and surrogate models that quantify their prediction uncertainties (i.e., probabilistic ML models such as GPs) will support their efficient and accurate execution by leveraging active learning principles; and (3) probabilistic ML and UQ capabilities could be leveraged by MOOSE's extensive user base.

Probabilistic ML deals with the development of surrogate models that can quantify complex multiphysics model prediction uncertainties. UQ deals with all aspects of identifying and inversely quantifying different sources of uncertainties, then forward propagating them to the model predictions. Probabilistic ML and UQ go hand-in-hand, leading to efficient approaches for active learning, Bayesian inference, Bayesian optimization, etc. Among the existing software for performing various aspects of probabilistic ML and UQ are UQPy \cite{Tsapetis2023x}, CUQIPy \cite{Riis2024x}, MUQ \cite{Parno2021x}, and PyApprox \cite{Jakeman2023x}, as discussed in \citet{Seelinger2024x}. Most of these software programs were written in Python. The development and demonstration of probabilistic ML and UQ capabilities presented herein is oriented toward the extensive user/developer community of MOOSE, which is written in C++. Moreover, MOOSE inherently supports massive parallelism, meaning that the probabilistic ML and UQ approaches can be scaled to use thousands of processors, thus leading to high levels of efficiency when dealing with complex multiphysics models. Ultimately, the right software tools can significantly enhance various stages of the research, development, and deployment processes for energy solutions, with different tools being better suited to specific scenarios. 

Massively parallel probabilistic ML and UQ in MOOSE is achieved through its \texttt{Sampler}, \texttt{MultiApp}, \texttt{Reporter}, and \texttt{Surrogate} systems. \texttt{Sampler} proposes new input parameter samples from the underlying probability distributions, \texttt{MultiApp} facilitates evaluation of the MOOSE computational model while handling massive parallelism, \texttt{Reporter} facilitates post-model-evaluation decision making, and \texttt{Surrogate} handles the training, evaluation, and retraining of probabilistic surrogates. These systems and their interaction are key to the development of GP variants, active learning, Bayesian inverse UQ, adaptive forward UQ, Bayesian optimization, evolutionary optimization, and Markov chain Monte Carlo (MCMC) in MOOSE. The modularity offered by these systems enables development of a multitude of probabilistic ML and UQ algorithms. These aspects will be discussed in detail later in this paper. Besides discussing the software implementation, this paper also demonstrates its application to five different types of computational problems: (1) Bayesian inverse UQ of fission product release from nuclear fuel, using parallel active learning; (2) very rare events analysis of a heat pipe (HP) nuclear microreactor, using active learning; (3) acceleration of advanced manufacturing process simulations, using multi-output GPs (MOGPs) and dimensionality reduction; (4) prediction of lid-driven cavity flow, using with deep GPs (DGPs); and (5) model parameter optimization of tritium diffusion for nuclear fusion, using batch Bayesian optimization.

This paper is organized as follows. Section \ref{sec:review} provides a theoretical review of the active learning, Bayesian inverse UQ, adaptive forward UQ, Bayesian optimization, evolutionary optimization, and MCMC methods relevant to MOOSE. Section \ref{sec:moose-code} details the MOOSE code implementations. Section \ref{sec:applications} discusses the impact to the five aforementioned energy applications. Lastly, Section \ref{sec:conc} summarizes the paper and presents the conclusions.

\section{Methodology Overview}\label{sec:review}

This section provides a theoretical overview of the probabilistic ML and UQ methods relevant to the MOOSE implementation.

\begin{figure}[htbp!]
\centering
\includegraphics[scale=0.45]{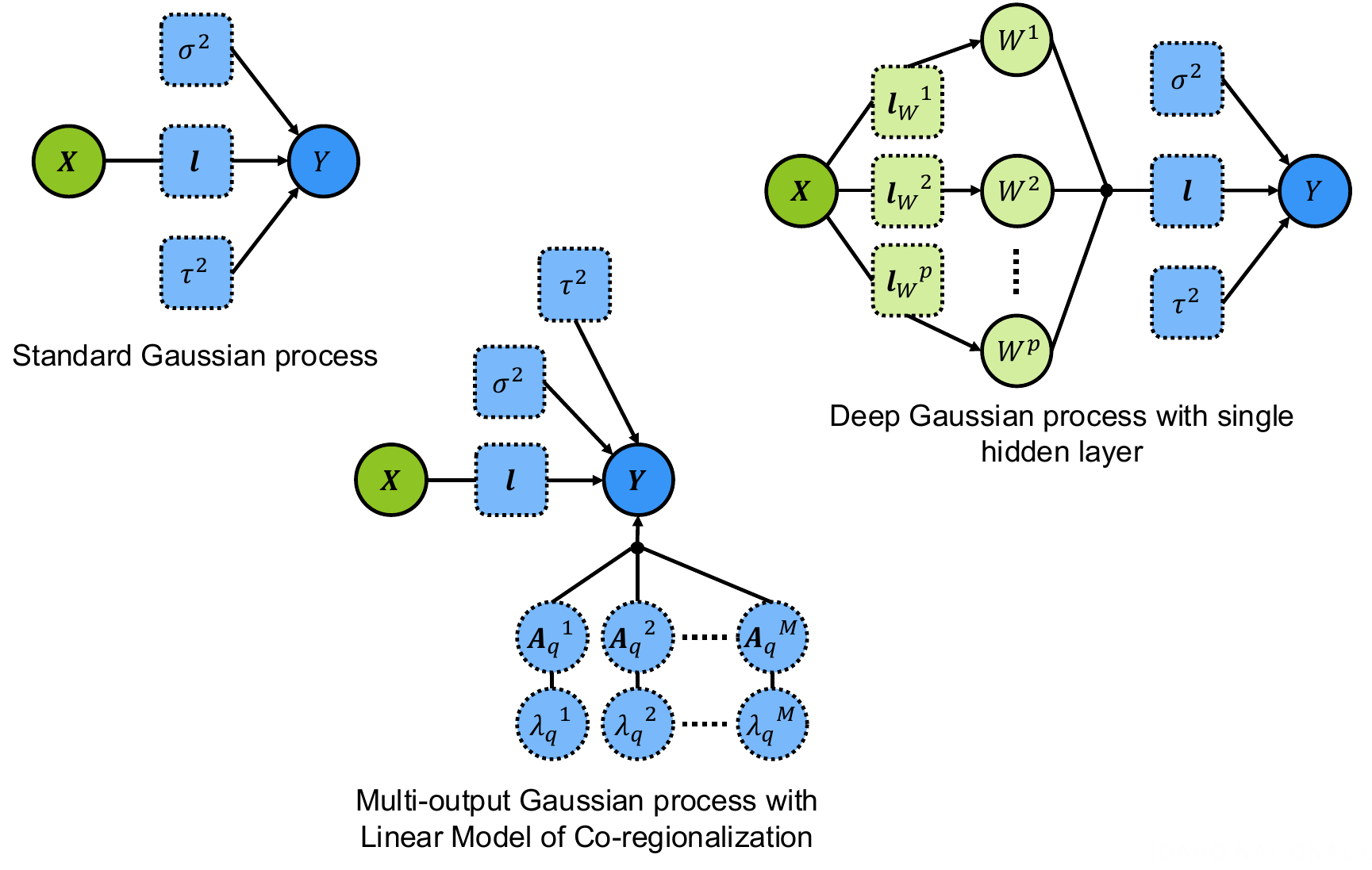} 
\caption{Graphical representation of the input ($\pmb{X}$) and output ($\pmb{Y}$) mapping of the three GP variants in MOOSE: standard GP, MOGP, and DGP. $\sigma^2$, $\pmb{l}$, and $\tau^2$, respectively, represent the amplitude scale, length scales, and noise variance hyperparameters. $\pmb{A}_q^i$ and $\lambda_q^i$ are the additional hyperparameters for an MOGP, and $\pmb{l}_W^i$ is the additional hyperparameters for a DGP. In MOOSE, these GP variants can be trained via either adaptive moment estimation (Adam) optimization (gradient-based) or MCMC sampling (gradient-free). Here, ``gradients" refers to gradients of the log-likelihood objective function.}
\label{fig:GP_STM}
\end{figure}

\subsection{Gaussian process variants}\label{sec:GP_all}

Figure \ref{fig:GP_STM} presents a graphical representation of the different GP variants in MOOSE. The theoretical details are briefly discussed below. The GP capabilities are used for Bayesian analysis of fission product release in an advanced nuclear fuel (Section \ref{sec:triso}), rare events analysis of a nuclear reactor (Section \ref{sec:heatpipe}), advanced manufacturing process modeling (Section \ref{sec:AM}), predicting fluid flow (Section \ref{sec:fluid}), and the optimization of a computational model for nuclear fusion (Section \ref{sec:fusion}), as discussed later in this paper.

\subsubsection{Standard Gaussian process}\label{sec:GP}


A standard GP is a stochastic process in which any finite collection of random variables follows a Gaussian distribution. Essentially, a GP describes a probability distribution over a function space and is discretized at certain points in the input space. A zero-mean GP is described as \cite{Williams2006a}: 

\begin{equation}
    \label{eqn:GP_prior}
    \pmb{y} \sim \mathcal{N}\bigg(\pmb{0},~k\big(\pmb{X},\pmb{X}^\prime\big)\bigg)
\end{equation}

\noindent where $\pmb{y}$ is the output vector of size $N$, $k(.,.)$ is the covariance function, and $\pmb{X}$ is the input matrix of size $N \times D$ ($D$ being the dimensionality of the inputs). As shown in Figure \ref{fig:GP_STM}, given input vectors $\pmb{x}$ and $\pmb{x}^\prime$, the scalar kernel function is described as:

\begin{equation}
    \label{eqn:gp_kernel}
    k(\pmb{x},\pmb{x}^\prime) = \sigma^2~\exp \Bigg(-\frac{1}{2}\sum_{d=1}^D\frac{(x_d-{x}_{d}^\prime)^2}{l_d^2}+ \tau^2 \mathbb{1}_{\pmb{x}=\pmb{x}^\prime} \Bigg)
\end{equation}

\noindent where $\pmb{l} = (l_1,\cdots l_D)$ is the vector of length scales, $\sigma^2$ is the amplitude, and $\tau^2$ is the noise term. When each input dimension is associated with its own length scale, the GP fitting procedure is referred to as automatic relevance determination (ARD) \cite{Williams2006a}, which is often used to implicitly determine the relevance of input variables. Note that $\pmb{x}$ is an input vector and $\pmb{X}$ is the input matrix at $N$ points. As such, $k(\pmb{x},\pmb{x}^\prime)$ is a scalar kernel function and $k(\pmb{X},\pmb{X}^\prime)$ is a covariance matrix of size $N \times N$. The parameters $\{\pmb{l},\sigma^2,\tau^2\}$ are the hyperparameters to be optimized by maximizing the log-likelihood function:

\begin{equation}
    \label{eqn:GP_likelihood}
    \ln~p(\pmb{y}~|~\pmb{X},\sigma^2,\pmb{l},\tau^2) \propto -\frac{1}{2}~\ln |k(\pmb{X},\pmb{X})| - \frac{1}{2}~\pmb{y}^T~k(\pmb{X},\pmb{X})^{-1}~\pmb{y}
\end{equation}

\noindent where $\pmb{X}$ and $\pmb{y}$ are the training inputs and outputs, respectively. Upon optimizing the hyperparameters, as discussed in Section \ref{sec:hyperopt}, the predictions of the GP on testing inputs $\pmb{X}_*$ constitute a Gaussian distribution:

\begin{equation}
    \label{eqn:GP_posterior}
    \begin{aligned}
    p(\pmb{y}_*~|~\pmb{X}, \pmb{X}_*, \pmb{y}) \sim \mathcal{N}\Big(~ & k(\pmb{X}_*,\pmb{X})~k(\pmb{X},\pmb{X})^{-1}~\pmb{y}, \\
    & k(\pmb{X}_*,\pmb{X}_*) - k(\pmb{X}_*,\pmb{X})~k(\pmb{X},\pmb{X})^{-1}~k(\pmb{X},\pmb{X}_*)~\Big)
    \end{aligned}
\end{equation}

\noindent where $p(\pmb{y}_*~|.)$ is the probabilistic prediction of the GP with mean vector $k(\pmb{X}_*,\pmb{X})~k(\pmb{X},\pmb{X})^{-1}~\pmb{y}$ and covariance $k(\pmb{X}_*,\pmb{X}_*) - k(\pmb{X}_*,\pmb{X})~k(\pmb{X},\pmb{X})^{-1}~k(\pmb{X},\pmb{X}_*)$. 

\subsubsection{Multi-output Gaussian processes (MOGP)}\label{sec:MOGP}

MOGPs model and predict vector outputs of size $M$. For any input matrix $\pmb{X}$, let the matrix of outputs be denoted by $\bar{\pmb{Y}} = [\pmb{y}_1,~\pmb{y}_2,\dots,~\pmb{y}_N]^\intercal$. Note that $\pmb{y}_i$ is of size $M\times 1$ and $\bar{\pmb{Y}}$ is of size $N\times M$. The matrix $\bar{\pmb{Y}}$ is vectorized and represented as $\hat{\pmb{y}}$ with size $NM \times 1$. $\hat{\pmb{y}}$ is modeled with a zero-mean Gaussian distribution prior, defined as:




\begin{equation}
    \label{eqn:mogp_2}
    \hat{\pmb{y}} \sim \mathcal{N}\Big(\hat{\pmb{0}},~\bar{\pmb{K}}\Big)
\end{equation}

\noindent where $\hat{\pmb{0}}$ is the mean vector and $\bar{\pmb{K}}$ is the full covariance matrix. $\bar{\pmb{K}}$ captures covariances across the input variables and the vector of outputs, and thus has a size of $NM \times NM$. $\bar{\pmb{K}}$ can be modeled in several different ways, as discussed in \cite{liu2018gp,alvarez2012gp}. As shown in Figure \ref{fig:GP_STM}, we will follow the linear model of co-regionalization (LMC), which distinctly models the covariances between the $N$ inputs and the $M$ outputs. Mathematically, the LMC is defined as \cite{liu2018gp,cheng2020gp}:

\begin{equation}
    \label{eqn:mogp_3}
    \bar{\pmb{K}} = \sum_{q=1}^Q \bar{\pmb{B}}_q \otimes \pmb{K}_q
\end{equation}

\noindent where $q$ denotes the basis index, $\bar{\pmb{B}}_q$ is the outputs covariance matrix of size $M \times M$ for the $q^{\textrm{th}}$ covariate, $\pmb{K}_q$ is the inputs covariance matrix of size $N \times N$ for the $q^{\textrm{th}}$ covariate, $Q$ is the total number of bases, and $\otimes$ denotes the Kronecker product. $\bar{\pmb{B}}_q$ is further defined as the sum of two matrices of weight \cite{cheng2020gp}:

\begin{equation}
    \label{eqn:mogp_4}
    \bar{\pmb{B}}_q = \pmb{A}_q \pmb{A}_q^\intercal + \textrm{diag}\Big(\pmb{\lambda}_q\Big)
\end{equation}

\noindent where $\pmb{A}_q$ and $\pmb{\lambda}_q$ are, respectively, the matrix (size $M\times R$) and vector (size $M\times 1$) of hyperparameters, both for the $q^{\textrm{th}}$ basis. The size $R$ is user defined and can be greater than or equal to 1. The larger the $R$, the more sophisticated the MOGP in modeling complex outputs. Furthermore, the size of $Q$ can also be greater than or equal to 1. Again, the larger the $Q$, the more sophisticated the MOGP in modeling complex outputs. In total, the MOGP with the LMC output covariance and the squared exponential input covariance kernel will have $Q~(D+1)~(M+1)~R$ hyperparameters to be optimized arising from $Q$ basis. If $Q=1$, the LMC reduces to the intrinsic co-regionalization model, with $(D+1)~(M+1)~R$ hyperparameters to be optimized. The MOGP log-likelihood function has a form similar to that of a scalar GP:

\begin{equation}
    \label{eqn:mogp_opt1}
    \mathcal{L} = -\frac{1}{2}~\ln |\bar{\pmb{K}}| - \frac{1}{2}~\hat{\pmb{y}}^T~\bar{\pmb{K}}^{-1}~\hat{\pmb{y}}-\frac{1}{2}~N~\ln(2\pi)
\end{equation}

\noindent Once the MOGP hyperparameters are optimized, as discussed in Section \ref{sec:hyperopt}, probabilistic predictions of the vector quantities of interest can be made. Given a prediction input $\pmb{x}_*$, the probability distribution of the vector outputs is given by:

\begin{equation}
    \label{eqn:mogp_pred1}
    p(\hat{\pmb{y}}_*|\pmb{x}_*,\hat{\pmb{y}},\bar{\pmb{x}},\pmb{\theta}) = \mathcal{N}(\hat{\pmb{\mu}}_*,~\bar{\pmb{\Sigma}}_{*})
\end{equation}

\noindent where $\bar{\pmb{x}}$ is the matrix of training inputs, $\hat{\pmb{\mu}}_*$ is the mean vector, and $\bar{\pmb{\Sigma}}_{*}$ is the covariance matrix. The mean vector is defined as:

\begin{equation}
    \label{eqn:mogp_pred2}
    \hat{\pmb{\mu}}_* = \bar{\pmb{K}}_{\hat{\pmb{y}}_*,\hat{\pmb{y}}} ~(\bar{\pmb{K}}_{\hat{\pmb{y}},\hat{\pmb{y}}})^{-1}~\hat{\pmb{y}}
\end{equation}

\noindent where $\bar{\pmb{K}}_{\hat{\pmb{y}}_*,\hat{\pmb{y}}}$ is the full covariance matrix of the training inputs and prediction inputs, and $\bar{\pmb{K}}_{\hat{\pmb{y}},\hat{\pmb{y}}}$ is the full covariance matrix of the training inputs. The covariance matrix $\bar{\pmb{\Sigma}}_{*}$ is defined as:

\begin{equation}
    \label{eqn:mogp_pred3}
    \bar{\pmb{\Sigma}}_{*} = \bar{\pmb{K}}_{\hat{\pmb{y}}_*,\hat{\pmb{y}}_*} - \bar{\pmb{K}}_{\hat{\pmb{y}}_*,\hat{\pmb{y}}}~(\bar{\pmb{K}}_{\hat{\pmb{y}},\hat{\pmb{y}}})^{-1}~\bar{\pmb{K}}_{\hat{\pmb{y}}_*,\hat{\pmb{y}}}^\intercal
\end{equation}

\noindent where $\bar{\pmb{K}}_{\hat{\pmb{y}}_*,\hat{\pmb{y}}_*}$ is the full covariance matrix of the prediction inputs.

\subsubsection{Deep Gaussian process}\label{sec:DGP}

Standard GPs entail the stationarity assumption, potentially limiting the GP's predictive performance (e.g., under regime changes in the input/output space). A stationary GP implies that the covariance between any two points depends only on the distance between them, not on their absolute locations. A DGP was first introduced by \citet{Damianou2013d,Damianou2015a} as a means of overcoming this stationarity assumption. By moving the inputs through hidden Gaussian layers, a DGP achieves non-stationarity even while using standard kernel functions (e.g., a squared exponential kernel) \cite{Sauer2023a}. Several DGP variants were proposed based on the optimization procedures used for determining the hyperparameters \cite{Salimbeni2017a,Dai2014a}. Herein, we rely on the DGP formulation of \citet{Sauer2023a}, who used MCMC for hyperparameter optimization. Considering a single-hidden-layer DGP (see Figure \ref{fig:GP_STM}), output $y$ is modeled as GPs over the hidden layer latents $\pmb{w}$, which are themselves modeled as a GP over the input $\pmb{x}$. The prior is mathematically described as:

\begin{equation}
    \label{eqn:dgp_1}
    \begin{aligned}
    y|\pmb{w} \sim \mathcal{N}\bigg(0,~k\big(\pmb{w},\pmb{w}^\prime\big)\bigg)\\
    \pmb{w} \sim \mathcal{N}\bigg(\pmb{0},~k\big(\pmb{x},\pmb{x}^\prime\big)\bigg)\\
    \end{aligned}
\end{equation}

\noindent Note that, for convenience, the prior is described for a scalar value of the output $y$ corresponding to the input vector $\pmb{x}$. In this case, the latents $\pmb{w}$ are a vector of size $p$. \citet{Sauer2023a} recommends that $p$ be equal to the size of the input vector. The log-likelihood function is the summation of log-likelihoods describing the mapping from $y$ to $\pmb{w}$ and from $\pmb{w}$ to $\pmb{x}$. Given $N$ training inputs, $\pmb{X}$, $\pmb{y}$, and $\pmb{W}$ have sizes of $N \times D$, $N$, and $N \times p$, respectively. $\pmb{W}^i$ is the vector of latents for the $i^{\textrm{th}}$ node in the hidden layer, and has dimensionality $N$. The compound log-likelihood function is given by:

\begin{equation}
    \label{eqn:dgp_2}
    \begin{aligned}
    &\ln~p(\pmb{y}~|~\pmb{W},\sigma^2,\pmb{l},\tau^2) \propto -\frac{1}{2}~\ln |k(\pmb{W},\pmb{W})| - \frac{1}{2}~\pmb{y}^T~k(\pmb{W},\pmb{W})^{-1}~\pmb{y}\\
    &\ln~p(\pmb{W}~|~\pmb{X},\pmb{l}_{W}) \propto \sum_{i=1}^{p} -\frac{1}{2}~\ln |k^i(\pmb{X},\pmb{X})| - \frac{1}{2}~(\pmb{W}^i)^T~k^i(\pmb{X},\pmb{X})^{-1}~\pmb{W}^i\\
    &\ln~p(\pmb{y}~|~\pmb{W},\sigma^2,~\pmb{X},\pmb{l},~\pmb{l}_{W},\tau^2) = \ln~p(\pmb{y}~|~\pmb{W},\sigma^2,\pmb{l},\tau^2) + \ln~p(\pmb{W}~|~\pmb{X},\pmb{l}_{W})\\
    \end{aligned}
\end{equation}

\noindent The DGP hyperparameters are optimized with respect to the log-likelihood function above, as discussed in Section \ref{sec:hyperopt}. For the testing inputs $\pmb{X}_*$, the latents are first predicted per:

\begin{equation}
    \label{eqn:dgp_3}
    \begin{aligned}
    &\mu_{w^i}(\pmb{X}_*) = k^i(\pmb{X}_*,\pmb{X})~k^i(\pmb{X},\pmb{X})^{-1}~\pmb{W}^i\\
    &\Sigma_{w^i}(\pmb{X}_*) = k^i(\pmb{X}_*,\pmb{X}_*) - k^i(\pmb{X}_*,\pmb{X})~k^i(\pmb{X},\pmb{X})^{-1}~k^i(\pmb{X},\pmb{X}_*)\\
    \end{aligned}
\end{equation}

\noindent Note that the index $i$ denotes the node in the hidden layer. Using these latents, the output mean and covariance matrix are predicted per:

\begin{equation}
    \label{eqn:dgp_4}
    \begin{aligned}
    &\pmb{\mu}_* = k(\pmb{W}_*,\pmb{W})~k(\pmb{W},\pmb{W})^{-1}~\pmb{y}\\
    &\pmb{\Sigma}_* = k(\pmb{W}_*,\pmb{W}_*) - k(\pmb{W}_*,\pmb{W})~k(\pmb{W},\pmb{W})^{-1}~k(\pmb{W},\pmb{W}_*)\\
    \end{aligned}
\end{equation}

\subsubsection{Gradient-based and gradient-free optimization methods for hyperparameter tuning}\label{sec:hyperopt}



For gradient-based optimization of the hyperparameters of the GP variants, MOOSE employs adaptive moment estimation (Adam) \cite{kingma2014adam}. Adam is a stochastic optimization algorithm that permits mini-batch sampling during the optimization iterations. In traditional Adam with regularization, the gradient update and hyperparameter update steps are defined as \cite{kingma2014adam}:  


\begin{equation}
    \label{eqn:mogp_opt2}
    \begin{aligned}
    &\pmb{g}_t \leftarrow \nabla \mathcal{L}_t(\pmb{\theta}_{t-1}) + \lambda~\pmb{\theta}_{t-1}\\
    &\pmb{\theta}_{t} \leftarrow \pmb{\theta}_{t-1} - \eta_t \Big(\alpha \hat{\pmb{m}}_t / (\sqrt{\hat{\pmb{\nu}}_t} + \varepsilon)\Big)\\
    \end{aligned}
\end{equation}

\noindent where $t$ is the iteration, $\pmb{\theta}$ represents the optimizable hyperparameters, $\pmb{g}$ is the gradient update, $\lambda$ is the regularization weight, $\alpha$ and $\varepsilon$ are internal parameters of the algorithm, $\hat{\pmb{m}}$ is the corrected first moment update, $\hat{\pmb{\nu}}$ is the corrected second moment update, and $\eta$ is the schedule multiplier. \citet{loshchilov2017adamw} proposed the AdamW algorithm, which modifies how the regularization is performed in Adam, thereby increasing its optimization performance. AdamW modifies the gradient update and hyperparameter update steps as follows \cite{loshchilov2017adamw}:

\begin{equation}
    \label{eqn:mogp_opt3}
    \begin{aligned}
    &\pmb{g}_t \leftarrow \nabla \mathcal{L}_t(\pmb{\theta}_{t-1})\\
    &\pmb{\theta}_{t} \leftarrow \pmb{\theta}_{t-1} - \eta_t \Big(\alpha \hat{\pmb{m}}_t / (\sqrt{\hat{\pmb{\nu}}_t} + \varepsilon) + \lambda~\pmb{\theta}_{t-1}\Big)\\
    \end{aligned}
\end{equation}
 
\noindent wherein we see that the regularization is decoupled from the gradient update step and instead added to the hyperparameter update step. \citet{loshchilov2017adamw} found that this decoupling generally enhanced the Adam algorithm's performance across the suite of case studies considered.

In MOOSE, gradient-free optimization is also available for tuning the GP hyperparameters, particularly the DGP. This is based on MCMC sampling via the elliptical slice sampler (ESS) and Metropolis-Hastings (MH) sampler. ESS is particularly well suited for fields $\pmb{f}$ with Gaussian priors $\mathcal{N}(\pmb{0},\pmb{\Sigma})$ \cite{Murray2010a}. A random angle $\gamma \sim \mathcal{U}(0,~2\pi)$ is drawn with the bounds set to $\gamma_{\textrm{min}} = \gamma - 2\pi$ and $\gamma_{\textrm{max}} = \gamma$. A new proposal for $\pmb{f}$ is then made with the acceptance rate $\alpha$, as shown below \cite{Murray2010a}:

\begin{equation}
    \label{eqn:ess_1}
    \begin{aligned}
    &\pmb{f}^* = \pmb{f}^{t-1} \cos{\gamma} + \pmb{f}^{\textrm{prior}} \sin{\gamma}\\
    &\alpha = \min{\Bigg(1,~\frac{\mathcal{L}(\pmb{f}^*)}{\mathcal{L}(\pmb{f}^{t-1})}\Bigg)}\\
    \end{aligned}
\end{equation}

\noindent where $t$ is the MCMC iteration index and $\mathcal{L}$ denotes the likelihood function. Crucially, in contrast to the MH sampler, if the proposal $\pmb{f}^*$ is rejected, the bounds on $\gamma$ are shrunken to $\gamma_{\textrm{min}} = \gamma ~(\textrm{if}~\gamma<0)$ and $\gamma_{\textrm{max}} = \gamma$ (O.W.). A new proposal for $\gamma$ is then made using $\mathcal{U}(\gamma_{\textrm{min}},~\gamma_{\textrm{max}})$. The procedure is repeated until the new proposal $\pmb{f}^*$ is accepted in the current iteration $t$. For DGPs in particular, \citet{Sauer2023a} proposed a hybrid version of ESS and the MH sampler in order to improve hyperparameter inference, and this version is implemented in MOOSE. At each MCMC iteration $t$, the MH sampler is first used to update the parameters $\pmb{l}$, $\sigma^2$, $\tau^2$, and $\pmb{l}_W^{i}$ in sequence, such as in a Gibbs sampling scheme. Then, by conditioning on these new values, the latents $\pmb{W}$ are updated using ESS. The updating for iteration $t$ is given by:

\begin{equation}
    \label{eqn:ess_2}
    \begin{aligned}
    \sigma^2[t],\tau^2[t]~&\textrm{via}~\texttt{MH}~\textrm{with}~p(\pmb{y}~|~\pmb{W},\sigma^2,\pmb{l},\tau^2)\\
    \pmb{l}[t]~&\textrm{via}~\texttt{MH}~\textrm{with}~p(\pmb{y}~|~\pmb{W},\sigma^2,\pmb{l},\tau^2)\\
    \pmb{l}_W^{i}[t]~&\textrm{via}~\texttt{MH}~\textrm{with}~p(\pmb{W}~|~\pmb{X},\pmb{l}_{W})~\forall~i \in \{1,\dots,p\}\\
    \pmb{W}^{i}[t]~&\textrm{via}~\texttt{ESS}~\textrm{with}~p(\pmb{y}~|~\pmb{W},\sigma^2,\pmb{l},\tau^2)~\forall~i \in \{1,\dots,p\}\\
    \end{aligned}
\end{equation}

\noindent Note that the combination of MH and ESS for updating at each MCMC iteration resembles a Gibbs sampling scheme. Also, $p(.)$ in Equation \eqref{eqn:ess_2} is used for decision making in either the MH sampler or ESS to accept/reject a proposed sample.

\subsection{Batch acquisition functions for parallelized active learning}\label{sec:AL}

MOOSE currently features several acquisition functions for a variety of tasks such as Bayesian optimization, Bayesian inverse UQ, and global surrogate fitting. These acquisition functions are dependent on the mean prediction $(\hat{\mu})$ and standard deviation $(\hat{\sigma})$ of the GP variant. Table \ref{tab:MOOSE_acq_functions} presents these acquisition functions and also lists their usage. Note that some of them have a tuning parameter $\lambda$ whose functionality depends on the usage. For example, $\lambda$ serves to boost either exploratory or exploitative behavior for Bayesian optimization and Bayesian inverse UQ tasks. In contrast, $\lambda$ is the failure threshold for a rare events analysis task. Also, for some GP variants such as MOGP, the mean prediction and standard deviation are vector quantities. In such a case, the computed acquisition function will also be a vector quantity that must be reduced to a scalar by using operations such as sum, average, maximum, minimum, or product.

\begin{table}[htb!]
\centering
\caption{Acquisition functions in MOOSE for active learning for tasks such as optimization, Bayesian inverse UQ, and global surrogate fitting.}
\label{tab:MOOSE_acq_functions}
\small
\begin{tabular}{ |c|c|c| }
\hline
\textbf{Acquisition function $a(\pmb{x})$} &  \textbf{Mathematical form} &  \textbf{Usage}\\
\hline
 Expected Improvement \cite{Chen2022xs} & $z\Phi(z/\hat{\sigma}) + \hat{\sigma} \phi(z/\hat{\sigma})$ & Bayesian optimization\\
\hline
Upper Confidence Bound \cite{Contal2013a} & $\lambda \hat{\sigma}+\hat{\mu}$ & Bayesian optimization\\
\hline
Probability of Improvement \cite{Chen2022xs} & $\Phi\big(\big(\hat{\mu}-\mathcal{M}(\pmb{x}^*)\big)/\hat{\sigma}\big)$ & Bayesian optimization\\
\hline
Bayesian posterior targeted \cite{ElGammal2023c} & $\exp(2 \lambda \hat{\mu}) \big(\exp(\hat{\sigma})- 1\big)$ & Bayesian inverse UQ\\
\hline
U-function \cite{Echard2011a,Dhulipala2022c} & $(\hat{\mu}-\lambda)/\hat{\sigma}$ & Rare events analysis\\
\hline
\Centerstack[c]{Expected Improvement \\ for Global Fit \cite{Lam2008a}}
  & $\big(\hat{\mu}-\mathcal{M}(\pmb{x}^*)\big)^2 + \hat{\sigma}^2$ & Global fitting\\
\hline
Coefficient of variation & $\hat{\sigma}/\hat{\mu}$ & Global fitting\\
\hline
\end{tabular}
\begin{tablenotes}
\small
\item[] $\phi:$ Gaussian probability density function (PDF), $\Phi:$ Gaussian cumulative distribution function (CDF), $\hat{\mu}:$ GP variant mean, $\hat{\sigma}:$ GP variant standard deviation, $\mathcal{M}:$ Computational model, $\pmb{x}^*:$ current best point, $\lambda:$ acquisition function parameter, and $z=\hat{\mu} - \lambda - \mathcal{M}(\pmb{x}^*)$
\end{tablenotes}
\end{table}
\normalsize

The acquisition functions listed in Table \ref{tab:MOOSE_acq_functions} permit sequential active learning, with one optimal location $\pmb{x}$ being specified to run the full-fidelity MOOSE model. However, sequential active learning can incur significant computational cost, as running the full-fidelity MOOSE model several times in sequence is expensive. To alleviate this, we used batch versions of the acquisition functions, where $b$ (a user-defined parameter) optimal locations of the inputs are specified to run the MOOSE model in parallel. For simplicity, we adopted the local penalization approach proposed by \citet{Zhan2017c}. In it, a correlation function between two inputs is first defined as:

\begin{equation}
    \label{eqn:al_5}
    Corr(\pmb{x}, \pmb{x}^\prime) = 1 - \exp\Big(\frac{-||(\pmb{x}- \pmb{x}^\prime) / \pmb{l}||}{2}\Big)
\end{equation}

\noindent where $\pmb{l}$ represents the length scales, as obtained through GP hyperparameter optimization. The $b$ optimal points for running the MOOSE model are defined as:

\begin{equation}
    \label{eqn:al_6}
    \begin{aligned}
        \pmb{x}^1 &= \argmax_{\pmb{x}} ~a(\pmb{x})\\
        \pmb{x}^2 &= \argmax_{\pmb{x}} ~a(\pmb{x})~Corr(\pmb{x}, \pmb{x}^1)\\
        \pmb{x}^b &= \argmax_{\pmb{x}} ~a(\pmb{x})~\prod_{i=1}^{b-1}Corr(\pmb{x}, \pmb{x}^i)\\
    \end{aligned}
\end{equation}

\noindent In this manner, we can select $b$ optimal points within each iteration of active learning by performing local penalization to mitigate any clustering of those points. These $b$ points can be evaluated in parallel by using a MOOSE model, and the GP variant is retrained by appending the input/output data with the new points. Other approaches for batch selection of the optimal points are also available, such as the Kriging Believer algorithm proposed by \citet{Ginsbourger2010a}. \citet{Wang2023s} provide a review of the recent developments in batch selection. These approaches will be pursued in MOOSE in the future. These active learning capabilities are used for Bayesian analysis of fission product release in an advanced nuclear fuel (Section \ref{sec:triso}), rare events analysis of a nuclear reactor (Section \ref{sec:heatpipe}), and optimizing a computational model in nuclear fusion (Section \ref{sec:fusion}), as discussed later in this paper.

\subsection{Inverse sampling and Bayesian inference}\label{sec:Bayes}

For inverse UQ, it is often of interest to calibrate computational models given the experimental data while quantifying the uncertainties associated with model parameters, model inadequacy (i.e., model structural error), and experimental noise. Following the Kennedy and O'Hagan framework \cite{Kennedy2001a}, the experimental data are defined to have originated from a generative model of the following form assuming independent and identically distributed experiments:

\begin{equation}
    \label{eqn:bt_5}
    \begin{aligned}
        &\mathcal{D}(\Theta_i) = \mathcal{M}(\pmb{\theta},~\Theta_i) + \delta (\Theta_i) + \varepsilon\\
        &\text{where,}~\varepsilon \sim \mathcal{L}(\sigma_\varepsilon)
    \end{aligned}
\end{equation}

\noindent where the $i^{\text{th}}$ experimental observation is indicated to be the model prediction plus a model inadequacy term $(\delta)$, plus a correction factor $(\varepsilon)$ to account for noise in the experimental data. In Equation \eqref{eqn:bt_5}, $\mathcal{M}$ is the computational model, $\pmb{\theta}$ are the model parameters, and $\Theta$ is the experimental configuration. The model inadequacy term is traditionally modeled with a standard GP, as further discussed in Section \ref{sec:GP}. The correction factor is treated as a random variable that follows a probability distribution generically defined as $\mathcal{L}$, and whose scale is $\sigma_\varepsilon$ and mean is 0. $\mathcal{L}$ is the likelihood function that evaluates the adequacy of the model predictions against the experimental data for a given $\pmb{\theta}$ and $\sigma_\varepsilon$:

\begin{equation}
    \label{eqn:bt_3}
    \mathcal{L}(\pmb{\theta}, \sigma_\varepsilon | \pmb{\Theta}, \mathcal{M}, \pmb{\mathcal{D}}) = \prod_{i=1}^N \mathcal{L}(\pmb{\theta}, \sigma_\varepsilon | \Theta_i, \mathcal{M}, \mathcal{D}_i)
\end{equation}

\noindent where the term within the product sign is specific to a given experimental configuration, and the product sign itself indicates that the experiments are independent and identically distributed. Specifically, under the Gaussian assumption, the likelihood function becomes:

\begin{equation}
    \label{eqn:bt_40}
    \mathcal{L}(\pmb{\theta}, \sigma_\varepsilon | \pmb{\Theta}, \mathcal{M}, \pmb{\mathcal{D}}) = \prod_{i=1}^N \mathcal{N}\big(\mathcal{D}(\Theta_i) - \mathcal{M}(\pmb{\theta},~\Theta_i) - \delta(\Theta_i),~\sigma_\varepsilon \big)
\end{equation}

\noindent With the likelihood function defined, the Bayesian inference problem entails quantifying the posterior distribution of $\{\pmb{\theta},\sigma_\varepsilon\}$ \cite{Kennedy2001a,Arendt2012a,Wu2018a,Radaideh2019a,Robbe2023a}:

\begin{equation}
    \label{eqn:bt_30}
    f(\pmb{\theta}, \sigma_\varepsilon | \pmb{\Theta}, \mathcal{M}, \pmb{\mathcal{D}}) \propto \mathcal{L}(\pmb{\theta}, \sigma_\varepsilon | \pmb{\Theta}, \mathcal{M}, \pmb{\mathcal{D}})~f(\pmb{\theta}, \sigma_\varepsilon)
\end{equation}

\noindent where $f(\pmb{\theta}, \sigma_\varepsilon)$ defines the prior distribution before observing new experimental data. The proportionality constant in Equation \eqref{eqn:bt_30} is a multidimensional integration over $\{\pmb{\theta},\sigma_\varepsilon\}$ and is typically unknown. Thus, MCMC techniques are traditionally used to solve the Bayesian inverse problem. 

\begin{figure}[htb!]
    \centering
    \begin{subfigure}[b]{0.35\textwidth} \centering
        \includegraphics[width=\linewidth]{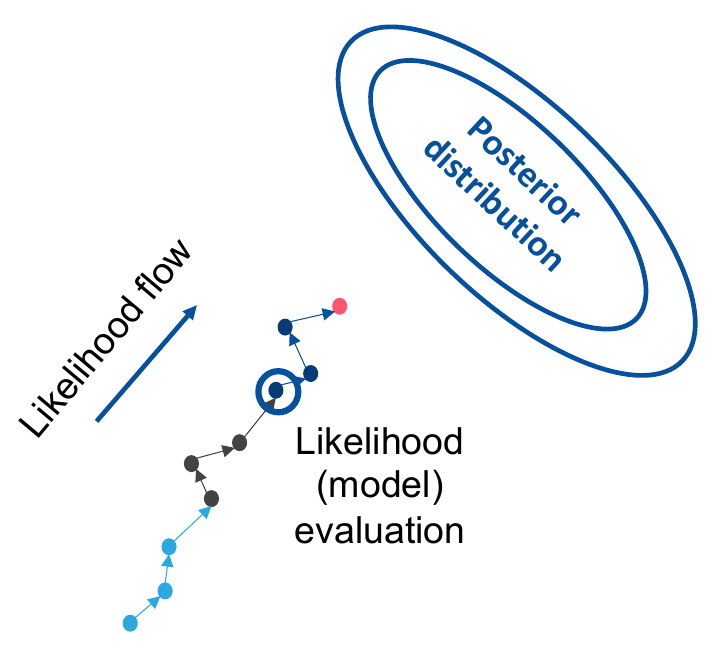}
        \caption{}
        \label{fig:serial_mcmc}
    \end{subfigure}
    \hfill
    \begin{subfigure}[b]{0.54\textwidth} \centering
        \includegraphics[width=\linewidth]{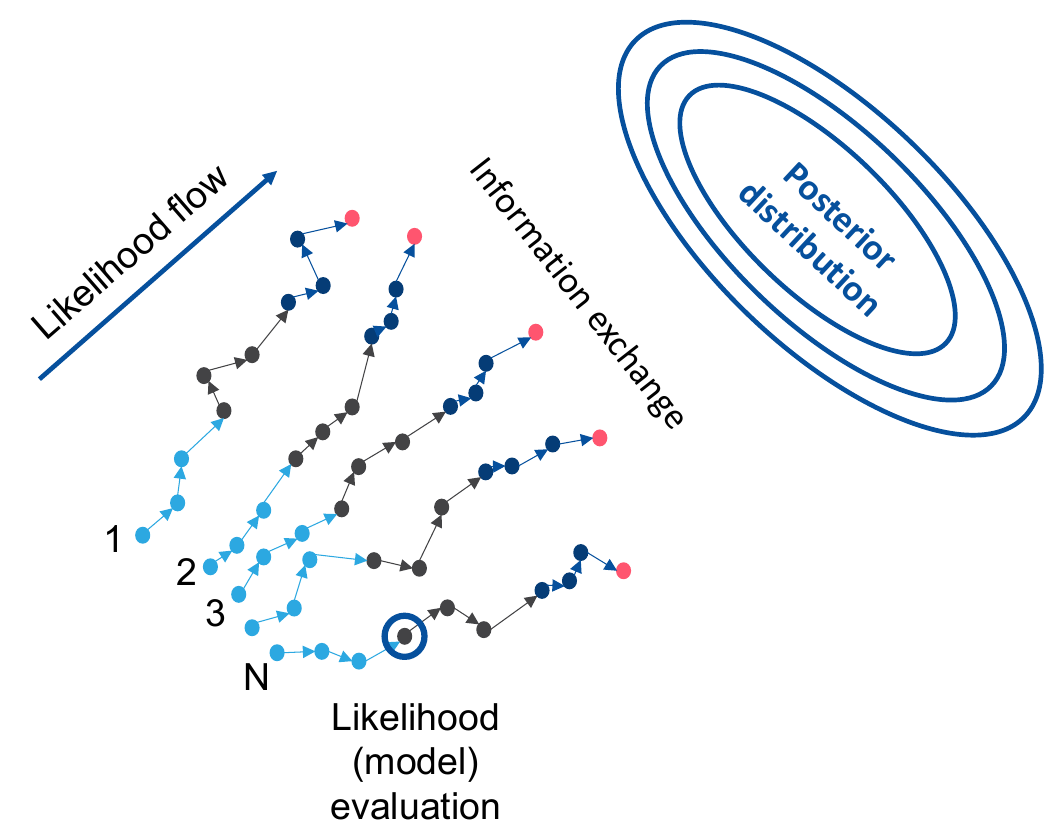}
        \caption{}
        \label{fig:parallel_mcmc}
    \end{subfigure}
    \caption{(a) Serial and (b) parallel/ensemble MCMC methods for obtaining samples from the posterior distribution. In comparison to serial MCMC samplers, parallel/ensemble MCMC samplers usually accelerate convergence to the posterior distribution.}
    \label{fig:serial_parallel_mcmc}
\end{figure}

MCMC techniques, widely regarded as the gold standard for solving the Bayesian inference problem, involve drawing samples from the posterior distribution described by Equation \eqref{eqn:bt_30}. Use of an MCMC sampler in practice is presented in Figure \ref{fig:serial_mcmc}. We start from an arbitrary realization of $\{\pmb{\theta},~\sigma\}$ and propose a new sample. The proposal can rely on the proposal distribution if the MCMC sampler falls under the MH class. Otherwise, it can be implicitly defined without requiring a proposal distribution, as in the case of an ensemble MCMC sampler \cite{Braak2006a,Goodman2010a}. In any case, the computational model is then evaluated for the newly proposed $\{\pmb{\theta},~\sigma\}$. Using the computational model output, the likelihood function is evaluated and the transition probability with respect to the old sample is computed. The new proposal is accepted with probability $t_{xy}$. Repeating the process of making a new proposal, evaluating the computational model and the likelihood function, and accepting/rejecting the proposal a sufficient number of times will give us the samples from the required posterior distribution.






This version of the MCMC sampler is serial in nature. Thus, it can take a significant number of serial steps to reach convergence, entailing many serial evaluations of the computational model. As this can be very expensive in practice, we will discuss parallelizable MCMC samplers that have multiple parallel Markov chains. Figure \ref{fig:parallel_mcmc} presents the working principle behind parallel MCMC samplers, which is similar to that of a serial MCMC sampler. At each step, $P$ parallel proposals are made, then the computational model corresponding to each proposal is evaluated. Since these model evaluations are independent of each other, they can be parallelized. The outputs are then used to compute the likelihood functions, and the Markov chains exchange information with each other to determine the next-best set of $P$ parallel proposals. The manner in which information exchange between chains is formulated differentiates the parallel MCMC samplers. Calderhead \cite{Calderhead2014a} proposed a parallelized version of the MH class of samplers. Goodman and Weare \cite{Goodman2010a} proposed a version of ensemble MCMC based on the affine invariance property, whereas Braak \cite{Braak2006a} proposed one based on differential evolution optimization \cite{Nelson2013a}. All these parallel MCMC variants are available in MOOSE. Interested readers are referred to \cite{Calderhead2014a,Goodman2010a,Braak2006a, Dhulipala2023uq} for the corresponding mathematical details. In addition to being massively parallelizable, parallel/ensemble samplers have been shown to accelerate convergence to the posterior, in comparison to the serial MCMC samplers. Studies such as Laloy and Vrugt \cite{Laloy2012a}, Foreman-Mackey et al. \cite{Foreman2019a}, and Opara and Arabas \cite{Opara2019a} discuss the convergence of MCMC samplers with the aid of metrics such as the Gelman-Rubin diagnostic \cite{vats2021mcmc} and the effective sample size.

For any new experimental configuration $\hat{\Theta}$, the posterior predictive distribution is:

\begin{equation}
    \label{eqn:bpp1}
        f\big(\mathcal{M}(\hat{\Theta},~\pmb{\theta})|\pmb{\Theta},\pmb{\mathcal{D}}\big) = \int_{\sigma_\varepsilon} \int_{\pmb{\theta}}\mathcal{L}(\pmb{\theta}, \sigma_\varepsilon | \pmb{\Theta}, \mathcal{M}, \pmb{\mathcal{D}})~f(\pmb{\theta}, \sigma_\varepsilon | \pmb{\Theta}, \mathcal{M}, \pmb{\mathcal{D}})~d\pmb{\theta}~d\sigma_\varepsilon
\end{equation}

\noindent where $\mathcal{L}(\pmb{\theta}, \sigma_\varepsilon | \pmb{\Theta}, \mathcal{M}, \pmb{\mathcal{D}})$ has the same form as in Equation \eqref{eqn:bt_3}. From the probability distribution of the model prediction described in Equation \eqref{eqn:bpp1}, statistics such as the median prediction and confidence bands can be inferred. This requires forward sampling techniques, discussed next. The inverse UQ capabilities are used for Bayesian analysis of fission product release in an advanced nuclear fuel (Section \ref{sec:triso}), as discussed later in this paper.

\subsection{Forward sampling}\label{sec:forwardUQ}

Forward sampling methods sample from a known probability distribution $q(\pmb{x})$. Traditional methods such as Monte Carlo sampling and Latin hypercube sampling (LHS) are available in MOOSE. When estimating certain statistics, Monte Carlo and LHS may require numerous evaluations of the model $\mathcal{M}$, thus becoming computationally intractable. There may also be cases in which directly drawing samples from the distribution $q(\pmb{x})$ is infeasible. Importance sampling addresses these concerns by sampling from an importance density $f(\pmb{x})$. The mean estimator of the quantity of interest $\mathcal{Q}\big(\mathcal{M}(\pmb{x})\big)$ is then computed via the modified equation \cite{Dhulipala2022cs}:

\begin{equation}
    \label{eqn:AIS_1}
    \hat{\mathcal{Q}} = \frac{1}{S}~\sum_{i=1}^S \mathcal{Q}\big(\mathcal{M}(\pmb{x}_i)\big)~\frac{q(\pmb{x}_i)}{f(\pmb{x}_i)}
\end{equation}

\noindent where $S$ is the number of samples drawn from the importance density $f(\pmb{x})$. The variance of the estimator is computed per \cite{Dhulipala2022cs}:

\begin{equation}
    \label{eqn:AIS_4}
    \textrm{Var}(\hat{\mathcal{Q}}) = \frac{1}{S}~\Bigg\{\frac{1}{S}~\sum_{i=1}^{S}\Bigg[\mathcal{Q}\big(\mathcal{M}(\pmb{x}_i)\big)~\frac{q(\pmb{x}_i)}{f(\pmb{x}_i)}\Bigg]^2-\hat{\mathcal{Q}}^2\Bigg\}
\end{equation}

\noindent A crucial component of importance sampling is the creation of importance density $f(\pmb{x})$. To estimate rare events, MCMC is a popular approach for creating $f(\pmb{x})$ by using an adaptive importance sampling scheme \cite{Au1999a,Zhao2015a,Zhang2019a}. For other applications, methods that use control variates \cite{Kawai2020a}, multilevel Monte Carlo \cite{Kebaier2018a}, and multifidelity modeling \cite{Peherstorfer2016a} have also been proposed to create $f(\pmb{x})$.



For more complex forward UQ applications such as global optimization and very rare events analysis, MOOSE also features a parallel subset simulation sampler \cite{Au2001a,Li2010a}. This is a variant of the sequential Monte Carlo sampler \cite{Bect2017cs}, with the goal being to sample from the failure or the optimal region. Subset simulation creates a series of intermediate thresholds---representing the suboptimal regions---that incrementally draw nearer to the optimal region. The method begins with regular Monte Carlo sampling for $N$ samples. The top $p_o \in [0,~1]$ samples are then selected in light of the quantity of interest $\mathcal{Q}\big(\mathcal{M}(\pmb{x})\big)$. Using these $p_o$ samples, Markov chains are initiated such that they propagate toward the optimal region and not in the other direction. If there are $N_M$ Markov chains, each is evaluated $\textrm{int}(N/N_M)$ times to obtain $N$ samples from this intermediate suboptimal region. The process of selecting the top $p_o$ samples from this intermediate region and initiating the Markov chains is repeated until convergence is achieved. As tens or hundreds of Markov chains are propagated in each subset, these and the corresponding MOOSE model evaluations can be massively parallelized. Note that parallelization can only be achieved across all the Markov chains, and not within the individual chains. More advanced versions of subset simulation have been proposed with respect to aspects such as the dynamic/adaptive intermediate thresholds \cite{Bect2017cs,Zhao2022cs} and the MCMC samplers \cite{Papaioannou2015a,Wang2019a,Shields2021a}. Building on the subset simulation sampler, other variants of this method---or of sequential Monte Carlo samplers in general---can be implemented in MOOSE at some point in the future. The forward UQ capabilities are used for rare events analysis of a nuclear reactor (Section \ref{sec:heatpipe}), as discussed later in this paper.

\subsection{Dimensionality reduction}\label{sec:dimRed}

MOOSE stochastic tools module supports linear principal component analysis (PCA), a dimensionality reduction technique widely used across multiple scientific disciplines \cite{wold1987principal}. Linear PCA can be used to determine a lower-dimensional space (latent space) that is closest to the given data in a discrete $L^2$ norm. Let $\pmb{s} \in \mathbb{R}^N$ be a high-dimensional vector ($N$ is large) representing the high-dimensional solution fields from numerical solvers in MOOSE. To discover a low-dimensional latent space by using PCA, we collect snapshots of the solution fields and organize them into a snapshot matrix $\pmb{S} = [\pmb{s}_1,\pmb{s}_2,...,\pmb{s}_{N_s}]$. For discrete problems such as the one presented here, singular value decomposition (SVD) is performed for a linear PCA analysis. Therefore, we can obtain the principal components of the snapshots (basis functions of the latent space) by computing the SVD of the snapshot matrix:

\begin{equation}
    \label{eq:svd}
    \pmb{S} = \pmb{U} \pmb{\Sigma} \pmb{V}^T
\end{equation}

\noindent where matrices $\pmb{U}$ and $\pmb{V}$ are unitary and contain the left and right singular vectors, respectively, whereas diagonal matrix $\pmb{\Sigma}$ contains the singular values. MOOSE relies on the parallel SVD solvers through the aid of SLEPc \cite{hernandez2005slepc}, enabling it to efficiently compress very high-dimensional output fields. The columns of $\pmb{U}$ are also called principal components, and can be used to approximate the high-dimensional snapshots per:

\begin{equation}
    \pmb{s} \approx \boldsymbol{U}_r \pmb{c}_r
    \label{eq:pod_inverse_mapping}
\end{equation}

\noindent where $\pmb{c}_r \in \mathbb{R}^r$ contains the expansion coefficients or coordinates in the lower-dimensional latent space, while matrix $\boldsymbol{U}_r$ contains the first $r$ principal components. The columns of $\boldsymbol{U}_r$ span the closest $r$-dimensional subspace to the snapshots in $\boldsymbol{S}$. Based on this expression and the fact that the principal components are orthonormal, we can map the snapshots to the latent space via the following operation:

\begin{equation}
    \pmb{c}_r = \boldsymbol{U}^T_r \pmb{s}
    \label{eq:pod_mapping}
\end{equation}

\noindent To determine the necessary number of principal components, (i.e., $r$) an explained variation-based approach is utilized that relies on the the singular values ($\sigma_{i}$) located on the diagonal of matrix $\boldsymbol{\Sigma}$:

\begin{equation}  
    r = \argmin_{1\leq r \leq N_s} \left(1 - \frac{\sum\limits_{i=1}^{r} \sigma^2_{i}}{\sum\limits_{i=1}^{N_s} \sigma_{i}^2}\right) < \tau
\label{eq:sv-decay}
\end{equation}

\noindent The above metric selects $r$ so that the relative sum of the squared singular values from $r$ to $N_s$ is lower than a given number $\tau\in(0,1]$. The dimensionality reduction capabilities are used advanced manufacturing process modeling (Section \ref{sec:AM}), as discussed later in this paper.

\section{MOOSE Code Implementations}\label{sec:moose-code}

\subsection{Background on the MOOSE Stochastic Tools Module}

The MOOSE stochastic tools module aims to efficiently and scalably sample parameters, run multiphysics models, and perform stochastic analyses, including UQ, sensitivity analysis, and surrogate model generation. In \citet{Slaughter2023a}, a more comprehensive and general overview of the module is presented. The following subsections describe the MOOSE systems relevant to the probabilistic ML and UQ techniques focused on in this paper.

\subsubsection{\texttt{Samplers} system}

The \texttt{Samplers} system represents a class of objects responsible for generating random samples. MOOSE provides a variety of objects for specific sampling strategies, including \texttt{MonteCarlo} and \texttt{LatinHypercube} for basic random sampling, \texttt{Quadrature} for sparse quadrature sampling, \texttt{AdaptiveImportance} and \texttt{ParallelSubsetSimulation} for MC-based forward-UQ sampling, and various objects for MC-based inverse-UQ sampling. For adaptive sampling schemes (e.g., MC-based sampling), these objects can gather data from associated objects so as to determine subsequent sets of samples---for instance, gathering whether or not a sample was rejected or accepted in the chain. \texttt{Samplers} also define how the multiphysics runs are parallelized. Typically, the number of parallel runs and the number of processors needed for each run are determined programmatically, though there are input parameters that allow for user control.

\subsubsection{\texttt{MultiApps} system}

\texttt{MultiApps} is a framework-level system in MOOSE that enables instantiation of independent simulations~\cite{gaston2015physics}. MOOSE utilizes this system to run multiphysics simulations during stochastic sampling and to gather the results. In particular, it leverages the flexibility in distributing simulations across processors, making the stochastic simulations both extremely scalable and memory efficient. Further details on the distribution of \texttt{MultiApps} for MOOSE are presented in \citet{Slaughter2023a}.

\subsubsection{\texttt{Reporters} system}

The MOOSE \texttt{Reporters} system provides an interface for declaring, manipulating, and gathering global data in a given application. MOOSE primarily utilizes this system to store data from \texttt{MultiApps} runs during the stochastic simulation. \texttt{Reporter} objects also handle heterogeneous storage of the data, keeping data distributed for memory efficiency and homogenizing them when necessary. \texttt{Reporters} is also the primary strategy for outputting data such as UQ results, typically in the form of JSON files.

\subsubsection{\texttt{Surrogates} system}

The \texttt{Surrogates} system in MOOSE provides the capability to train and evaluate meta-models. \texttt{Trainers} are responsible for gathering parameter values from \texttt{Samplers} and responses from \texttt{Reporters} to compute the necessary data for model generation. These data can be declared globally or output for later use. \texttt{Surrogates} then takes the trained model and provides an interface for evaluating it. Specified \texttt{Trainers} and \texttt{Surrogates} are accessible from any MOOSE object in order to either evaluate the model based on specific parameters or retrain them on-the-fly. All the GP variants are built using the \texttt{Surrogates} system.

\subsection{Modularity: understanding the \texttt{Sampler}, \texttt{MultiApp}, \texttt{Reporter}, and \texttt{Surrogate} interaction}

The \texttt{Sampler}, \texttt{MultiApp}, \texttt{Reporter}, and \texttt{Surrogate} systems in MOOSE afford extensive modularity and enable development of many variants of active learning, forward/inverse UQ, and Bayesian optimization algorithms. Moreover, these algorithms can be implemented in an inherently parallel manner by calling several instances of the computational MOOSE model in parallel, using the \texttt{MultiApp} system. Understanding how the \texttt{Sampler}, \texttt{MultiApp}, \texttt{Reporter}, and \texttt{Surrogate} systems interact with each other---as well as their order of execution within MOOSE---is key to implementing these algorithms. This section discusses the interaction between these systems.

\begin{figure}[htb!]
    \centering
        \includegraphics[width=1.0\linewidth]{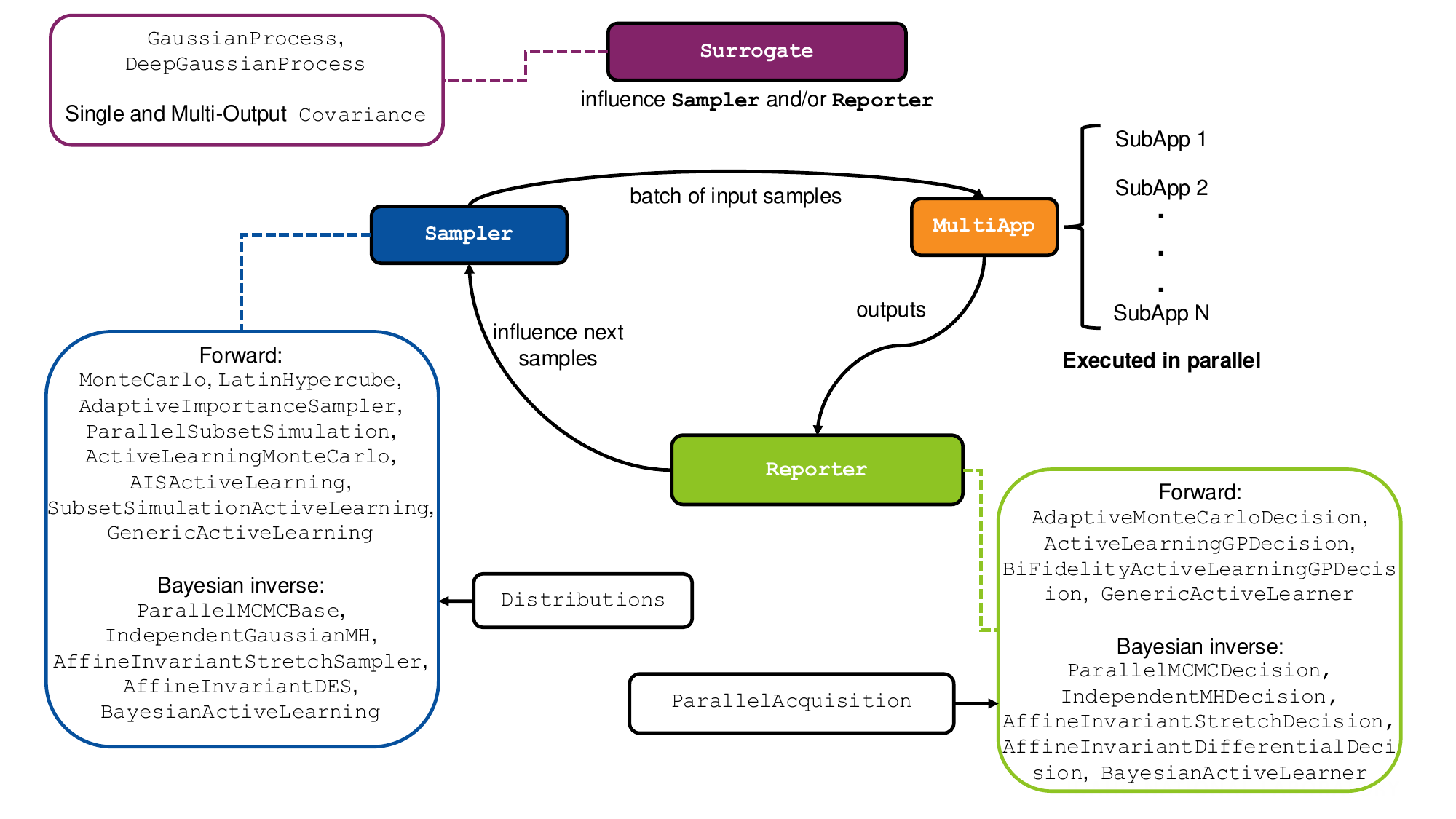}
    \caption{\texttt{Sampler}, \texttt{MultiApp}, \texttt{Reporter}, and \texttt{Surrogate} system interaction in MOOSE for performing parallel active learning. The available objects deriving off of \texttt{Sampler} and \texttt{Reporter} are also shown in regard to supporting tasks such as forward/inverse UQ, Bayesian optimization, and active learning with different GP variants.}
    \label{fig:MOOSEProbML}
\end{figure}

For the sake of simplicity, the interaction among \texttt{Sampler}, \texttt{MultiApp}, and \texttt{Reporter} is discussed first. \texttt{Sampler} proposes new samples from the underlying probability distributions, using objects in the \texttt{Distributions} system. These proposed samples are stored in a global array, with the rows containing the samples to be executed in parallel and the columns representing the number of parameters to the computational model. The numerical simulations corresponding to the  proposed samples are automatically executed in parallel, if the user desires, via the \texttt{MultiApp} system. Upon execution, the simulation outputs are received by the \texttt{Reporter} system and stored in a JSON file. Under simple schemes such as Monte Carlo or LHS, the \texttt{Reporter} system only outputs the simulation results and the \texttt{Sampler} system then moves on to propose the next batch of samples, without any influence from the previously proposed samples or their simulation outcomes. In schemes such as adaptive Monte Carlo and MCMC, the \texttt{Reporter} system plays a more crucial role of influencing the next batch of samples proposed by the \texttt{Sampler} system, depending upon the simulation outcomes of the previously proposed batch of samples. Several adaptive Monte Carlo and MCMC algorithms such as adaptive importance sampling and parallel subset simulation for forward UQ and parallel MH, and ensemble MCMC for inverse UQ, fit well within the \texttt{Sampler}, \texttt{MultiApp}, and \texttt{Reporter} interaction scheme in MOOSE. For Bayesian inverse UQ problems, the \texttt{Sampler} system performs the additional function of collecting the user-supplied experimental configuration data and combining them with the proposed samples of model parameters by creating combinations of these parameters and experimental configurations. Owing to the inherent parallelization via the \texttt{MultiApp} system, algorithms such as parallel subset simulation, parallel MH, and ensemble MCMC, which rely on multiple Markov chains, can be massively parallelized in terms of the computational model calls. Figure \ref{fig:MOOSEProbML} presents the \texttt{Sampler}, \texttt{MultiApp}, and \texttt{Reporter} system interaction flowchart, along with several objects available in MOOSE for forward and inverse UQ applications.

Next, we will discuss the \texttt{Surrogate} system's influence on the interaction among the \texttt{Sampler}, \texttt{MultiApp}, and \texttt{Reporter} systems. Training, evaluation, and active/online learning of the GP variants in MOOSE are handled by \texttt{Surrogate} and \texttt{Trainer}. The \texttt{Surrogate} system can be easily coupled to the \texttt{Reporter} system to influence its behavior and/or that of the \texttt{Sampler} system. For example, in parallel active learning tasks such as forward/inverse UQ and Bayesian optimization, the GP surrogate variant, based on its predictive uncertainties and the acquisition function values, tells the \texttt{Sampler} system the best sets of input parameters under which to call the MOOSE computational model during the next iteration. After evaluating the computational model, in parallel, the outputs will be obtained by the \texttt{Reporter} system, which retrains the GP variant with the appended new data. The \texttt{Reporter} system will then query the acquisition function about the next-best sets of input parameters, and this process repeats until reaching a user-specified number of outer iterations. \texttt{GaussianProcess} and \texttt{DeepGaussianProcess} surrogates are currently derivable off of the \texttt{Surrogate} system. Both rely on the \texttt{Covariance} system to set up the training data input/output covariances (output covariances are only required for the MOGP surrogate). They also rely on the \texttt{GaussianProcess} class, which handles the training and retraining by using the gradient-based Adam algorithm or gradient-free MCMC sampling. Here, ``gradients" refers to gradients of the log-likelihood function of the GP variant. Figure \ref{fig:MOOSEProbML} indicates how the \texttt{Surrogate} system influences the interaction among \texttt{Sampler}, \texttt{MultiApp}, and \texttt{Reporter}, and supports parallelized active learning. Moreover, a pre-trained GP surrogate variant saved as an .rd (restartable data) file can be loaded and evaluated by using a combination of user-specified \texttt{Sampler} and \texttt{Reporter} objects, without calling the MOOSE computational model.

\subsection{Example implementation of parallelized active learning}

\begin{figure}[htb!]
    \centering
    \begin{subfigure}[b]{1\textwidth} \centering
        \includegraphics[width=0.8\linewidth]{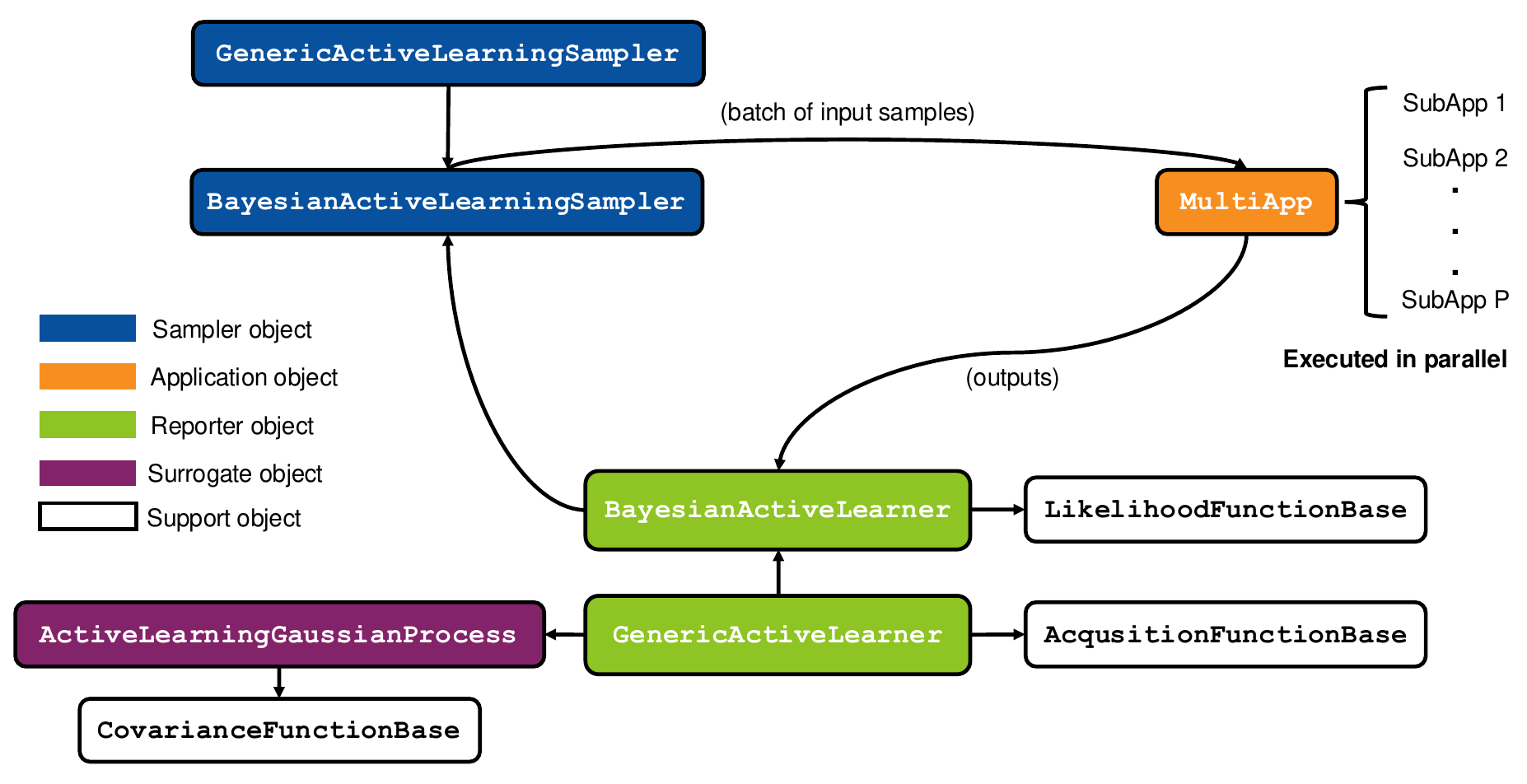}
    \caption{}
    \label{fig:parallel_al_example_1}
    \end{subfigure}
    \hfill
    \begin{subfigure}[b]{1\textwidth} \centering
        \includegraphics[width=0.5\linewidth]{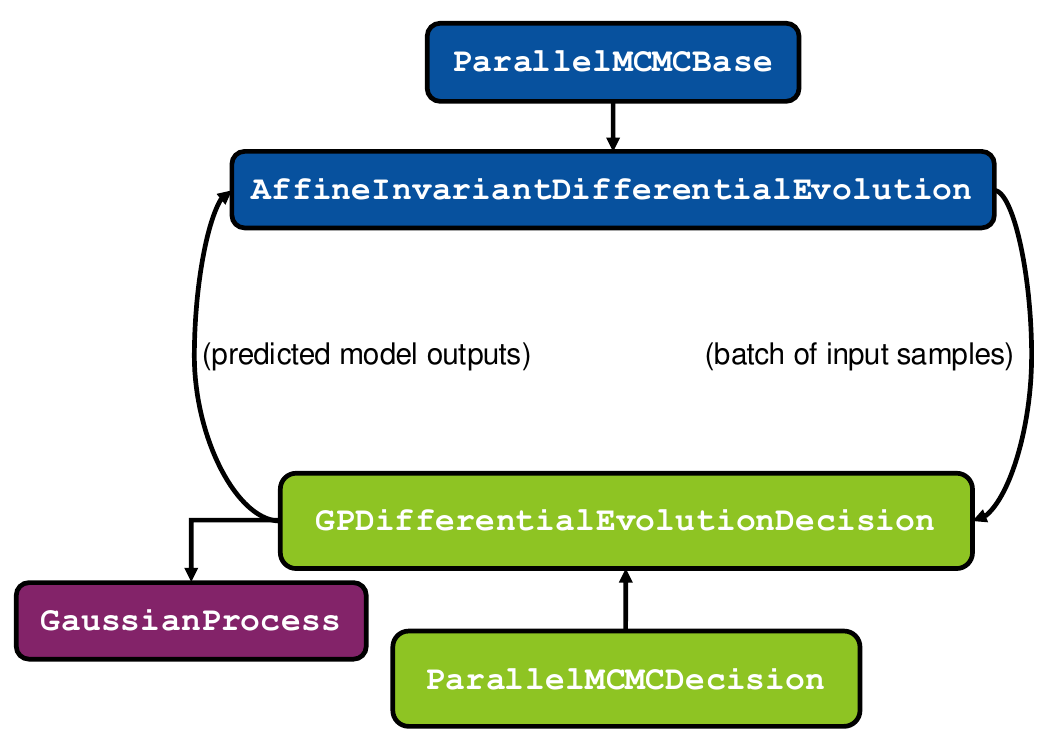}
        \caption{}
        \label{fig:parallel_al_example_2}
    \end{subfigure}
    \label{fig:parallel_al_example}
    \caption{(a) MOOSE objects and their dependencies for performing parallel active learning for Bayesian optimization and Bayesian UQ applications by leveraging the \texttt{Sampler}, \texttt{MultiApp}, \texttt{Reporter}, and \texttt{Surrogate} interaction. Note that the combination of \texttt{GenericActiveLearningSampler} and \texttt{GenericActiveLearner} performs Bayesian optimization. \texttt{BayesianActiveLearningSampler} and \texttt{BayesianActiveLearner} are derived objects for Bayesian UQ, and they consider the experimental configurations and likelihood functions supplied by the user. For Bayesian UQ, the actively trained GP prioritizes regions of high log-likelihood and is saved as an .rd file. (b) Evaluation phase of the actively trained GP for Bayesian UQ by leveraging the MCMC sampling objects; specifically, the \texttt{AffineInvariantDifferentialEvolution} for proposing new samples and the \texttt{GPDifferentialEvolutionDecision} for decision-making. Here, the GP model directly predicts the log-likelihood values under different input parameters and experimental configurations, thus circumventing evaluation of the computational MOOSE model.}
\end{figure}

An example implementation of parallel active learning capabilities in MOOSE---via leveraging the \texttt{Sampler}, \texttt{MultiApp}, \texttt{Reporter}, and \texttt{Surrogate} interaction---will now be discussed for Bayesian UQ and Bayesian optimization applications. Figure \ref{fig:parallel_al_example_1} presents the MOOSE objects and their dependencies. This schematic is comprised of the following main components:
\begin{itemize}
    \item \textbf{\texttt{GenericActiveLearningSampler}/\texttt{BayesianActiveLearningSampler}:} \newline \texttt{GenericActiveLearningSampler} creates a large population of input samples at each iteration, and this is retrieved by the \texttt{Reporter} object to facilitate optimization of the acquisition function. Importantly, this object also facilitates evaluation of the computational model via the \texttt{MultiApp} system for a best batch of inputs, as informed by the GP model. \texttt{BayesianActiveLearningSampler} derives from  \texttt{GenericActiveLearningSampler} and is tailored for Bayesian UQ applications such that it considers the experimental configurations. Specifically, before sending the inputs to the \texttt{MultiApp} system, \texttt{BayesianActiveLearningSampler} combines them with the experimental configurations.
    \item \textbf{\texttt{GenericActiveLearner}/\texttt{BayesianActiveLearner}:} \newline \texttt{GenericActiveLearner} optimizes the acquisition function via the \texttt{GaussianProcess} surrogate and selects the next-best set of inputs to the \texttt{Sampler} object. The acquisition function is optimized by selecting the best $P$ inputs from among the large population of samples created earlier in the iteration by the \texttt{GenericActiveLearningSampler}. \texttt{BayesianActiveLearner} derives from \texttt{GenericActiveLearner} to compute the log-likelihood function, which serves as the training/retraining data for the GP for Bayesian UQ applications.
    \item \textbf{Support objects:} \texttt{CovarianceFunctionBase} constructs covariances for the GP object, based on the kernel specified by the user. \texttt{LikelihoodFunctionBase} evaluates the likelihood function, given inputs and model outputs based on the user-specified distribution. \texttt{AcquisitionFunctionBase} computes the acquisition function specified by the user and performs local penalization when selecting the best $P$ input samples.
\end{itemize}

\texttt{GenericActiveLearningSampler} and \texttt{GenericActiveLearner} can readily perform batch Bayesian optimization for maximizing a user-defined objective evaluated via a MOOSE computational model. For Bayesian UQ, \texttt{BayesianActiveLearningSampler} and \texttt{BayesianActiveLearner} train a GP model by prioritizing regions of high log-likelihood via the Bayesian posterior targeted acquisition function detailed in Table \ref{tab:MOOSE_acq_functions}. The trained GP model is saved as an .rd file. This will be used in conjunction with MCMC objects such as the \texttt{AffineInvariantDifferentialEvolution} sampler \texttt{GPDifferentialEvolutionDecision} reporter for sampling from the posterior distribution. Doing so circumvents evaluation of the MOOSE computational model, since the trained GP model will directly predict the log-likelihood values during forward evaluation. The flowchart in Figure \ref{fig:parallel_al_example_2} details the use of an actively trained GP model for sampling from the posterior distribution.

\section{Application Demonstrations}\label{sec:applications}

\subsection{Parallel active learning for Bayesian inverse UQ of TRISO nuclear fuel fission product release}\label{sec:triso}

This section uses the active learning with GP and forward UQ capabilities discussed in Sections \ref{sec:AL} and \ref{sec:Bayes}, respectively.

Tristructural isotropic (TRISO) particle fuel is proposed for use in advanced reactors because of its high temperature resistance. Its protective layers are intended to encapsulate the fission products, which these reactor designs are based on. Thus, it critical to assess the predictive uncertainties in the TRISO fission product release model. To this end, inverse UQ of the TRISO fission product release model is necessary to quantify the uncertainties due to model parameters, model inadequacy, and experimental noise. A $25$-mm-long, $6$-mm-radius cylindrical fuel compact can contain approximately 10,000--15,000 TRISO particles, each with a radius of around 375--430 $\mu$m \cite{Petti2012a}. Each TRISO particle has several protective layers around the fuel kernel---namely, the buffer, inner pyrolitic carbon (PyC), silicon carbide, and outer PyC layers. Fission products, particularly silver release, are modeled using the BISON fuel performance code \cite{WilliamsonBISON,hales2021modeling}, which is a MOOSE-based application. The diffusion process of fission products in TRISO particles requires computation of the fuel temperature (if not prescribed), temperature-dependent diffusion coefficients, source rates for the fission products, and the particle geometry. Material models were developed in BISON for each type of material in the TRISO particles: the buffer, the PyC layers, the silicon carbide layer, and the fuel kernel. Fission product diffusion is governed by the Fickian diffusion equation, wherein the diffusivity of the fission products is in units of m$^2$/s, and is normally estimated via an effective diffusivity defined per an Arrhenius law. See \citet{WilliamsonBISON,hales2021modeling} for further details on the modeling using BISON. The values for the pre-exponential factor $D_i$ and activation energy $Q_i$ in the Arrhenius equation for the different TRISO layers are usually calibrated from existing experimental data. A sensitivity analysis conducted in \citet{Dhulipala2024z} concluded that the pre-exponential factors of the fuel kernel and PyC layer are, in comparison to the other model parameters, unimportant in predicting fractional silver release, which is the fission product of interest herein. Hence, the parameter space of interest is $\pmb{\theta} = \{Q_{kernel},~Q_{ipyc},~D_{sic},~Q_{sic}\}$ when considering the Arrhenius equation for silver diffusivity. Experimental datasets on the observed silver release from TRISO particles are available from the Department of Energy Advanced Gas Reactor program. This enables inverse calibration and UQ of the TRISO model parameter space. At the same time, it is also of interest to quantify the predictive uncertainty associated with model inadequacy and experimental noise. We used the massively parallel MCMC samplers and parallelizable active learning in MOOSE to inversely quantify the model parameters $\pmb{\theta}$ and the sigma term (model inadequacy plus experimental noise). Thanks to the Advanced Gas Reactor program, 32 experimental data points on the observed silver release have been made available, and were used for the inverse UQ process \cite{Stempien2021}.

The approaches to inversely assess the uncertainties in the model parameters and model inadequacy plus experimental noise are detailed below.

\begin{itemize}
    \item \textit{Parallel MCMC:} The TRISO fractional silver release predictions and observations are bounded between 0 and 1. So, we used a truncated normal likelihood function to assess the model predictions against the experimental data. The inversely calibrated parameters were $\{\pmb{\theta}, \sigma\}$, and the prior distributions for all the parameters were uniformly distributed. We used the differential evolution sampler \cite{Braak2006a} in MOOSE to inversely quantify the uncertainties in $\{\pmb{\theta}, \sigma\}$. For this purpose, we used 50 parallel chains, each executing the MOOSE model 32 times (i.e., the number of experimental data points), in parallel, to evaluate the likelihood function. As a result, $1,600~(\textrm{i.e.,}~50 \times 32)$  processors were employed to perform inverse UQ for a total of 500 serial iterations in the differential evolution sampler.
    \item \textit{Parallel active learning:} For this, we used the same likelihood formulation and priors as before. We used a standard GP to predict the fractional silver release of the MOOSE model. For active learning, we relied on the Bayesian posterior targeted acquisition function from Table \ref{tab:MOOSE_acq_functions} to actively acquire new training data by running the MOOSE model. We also combined this acquisition function with the local penalization approach (Equations \eqref{eqn:al_5}--\eqref{eqn:al_6}) to acquire a batch of new training data. We set the batch size to 10 and performed 80 serial iterations of active learning. At the end of the 80 iterations, we observed that a convergence metric had sufficiently stabilized. Then, using the actively trained standard GP, we performed differential evolution sampling, just as before, by replacing the MOOSE model evaluations. This led to an approximated posterior distribution of $\{\pmb{\theta}, \sigma\}$.
\end{itemize}

\begin{figure}[htb!]
    \centering
    \begin{subfigure}[b]{0.49\textwidth} \centering
        \includegraphics[width=0.85\linewidth,trim={0 0 6.5cm 0},clip]{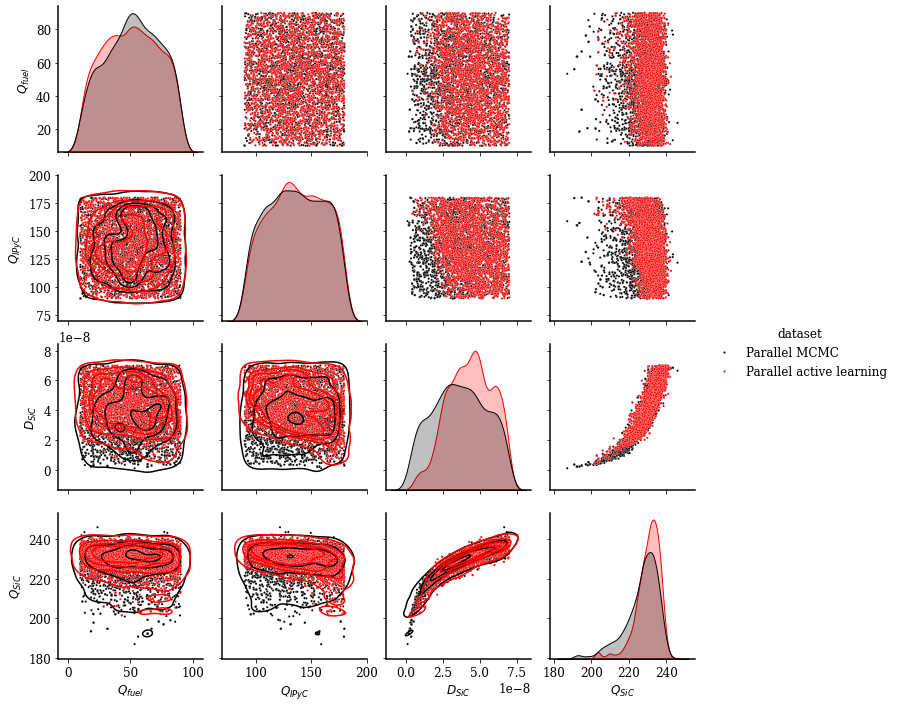}
    \caption{}
    \label{fig:al_params_triso}
    \end{subfigure}
    \hfill
    \begin{subfigure}[b]{0.49\textwidth} \centering
        \includegraphics[width=0.75\linewidth]{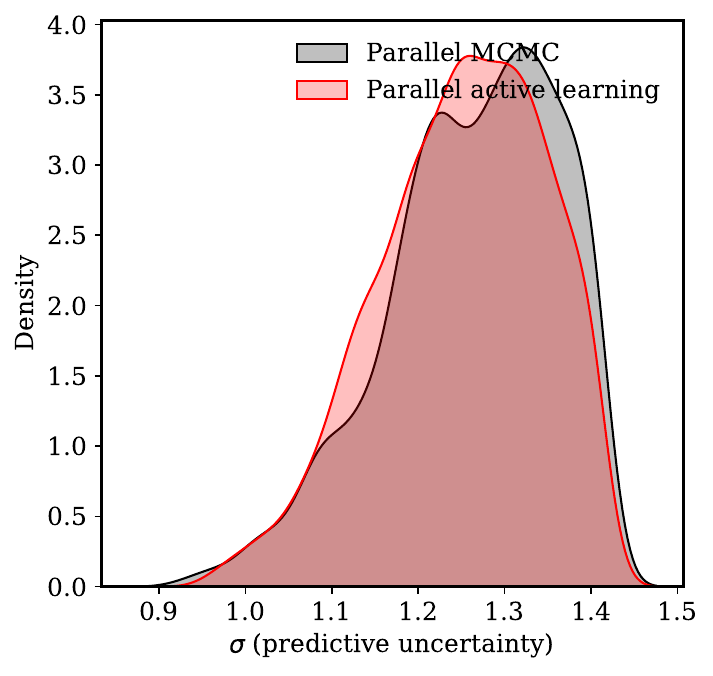}
        \caption{}
        \label{fig:al_sigma_triso}
    \end{subfigure}
    \label{fig:al_triso}
    \caption{Comparison of the posterior distributions of (a) the model parameters $\pmb{\theta}$ and (b) the sigma term (model inadequacy plus noise) in regard to parallel MCMC and parallel active learning approaches for the TRISO fuel silver release case.}
\end{figure}

Figure \ref{fig:al_params_triso} presents the inversely quantified posterior distributions of $\pmb{\theta}$, comparing the parallel MCMC and parallel active learning approaches. Note that, in general, parallel active learning gives posterior distributions consistent with parallel MCMC, which is considered to be the reference solution. Between the model parameters $D_{SiC}$ and $Q_{sic}$, we see a strong non-linear correlation, as shown in the subplots located in the third row, fourth column and the fourth row, third column. Parallel active learning is able to capture this non-linear correlation, though it struggles near the bottom left tip, where there is a small concentration of probability density. Figure \ref{fig:al_sigma_triso} presents the posterior distribution of the sigma $(\sigma)$ term, which captures the model inadequacy plus the experimental noise. Again, parallel MCMC and parallel active learning produce highly consistent results.   

\begin{figure}[htb!]
    \centering
        \includegraphics[width=0.4\linewidth]{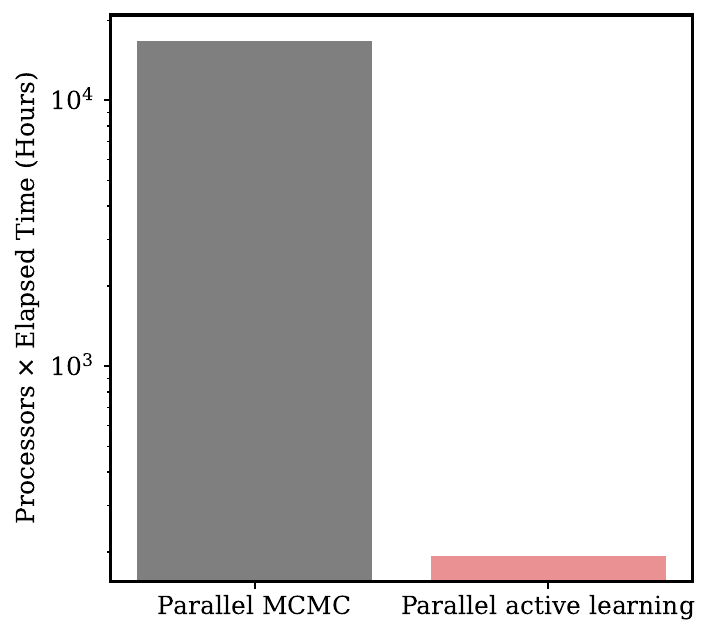}
        \caption{Comparison of the computational cost of performing inverse UQ in regard to the parallel active learning and parallel MCMC approaches for the TRISO fuel silver release case. (Computational cost is measured as the product of the number of processors required times the elapsed time necessary to solve the inverse UQ problem.)}
    \label{fig:al_cost_triso}
\end{figure}

Figure \ref{fig:al_cost_triso} compares the computational cost of inverse UQ in regard to parallel active learning and parallel MCMC. Computational cost is measured as the product of the number of processors required times the elapsed time necessary to solve the inverse UQ problem. Parallel active learning has shown to have a computational cost at least three orders of magnitude smaller than parallel MCMC, which is considered the reference solution, while still delivering satisfactory posterior uncertainties. Capturing features in the posterior distribution like sharp tails can be accomplished by increasing the number of iteration or using a better acquisition function.

\subsection{Active learning variance reduction for very rare events analysis of a heat-pipe nuclear microreactor}\label{sec:heatpipe}

This section uses the active learning with GP and forward UQ capabilities discussed in Sections \ref{sec:AL} and \ref{sec:forwardUQ}, respectively.

This section demonstrates the use of MOOSE ProbML capabilities for estimating very rare events, based on an HP nuclear microreactor model. Very rare events correspond to low failure probabilities on the order of $10^{-6}$ or lower. Unlike other types of nuclear reactors, HP-cooled microreactors must consider additional failure modes stemming from heat transfer limitations governing HP operability. These bounding limits constrain how much heat can be removed by the HPs, depending mainly on the HP design parameters and its temperature. Failure limits are computed using the MOOSE-based application Sockeye \cite{sockeye}, based on the design parameters specified in \citet{Terlizzi2023}, with the pore radius increased to 45~$\mu$m (to lower the capillary limit). Even though the sonic and viscous limits are not catastrophic---in that the HPs can recover after reaching them---for the purpose of this demonstration, all these limits are considered when determining failure probability. As manufacturing and thermal property uncertainties are very much design specific, and because the model considered herein is a prototypical design, this demonstration only serves as a proof-of-concept of MOOSE's methodological implementations for computing very low failure probabilities. As such, the reported values of the failure probabilities should not be directly applied to assess the safety of HP reactors.

The MOOSE computational model consists of a single HP. It employs the effective heat conduction model in Sockeye \cite{sockeye}, with the HP vapor core being represented as a material with extremely high thermal conductivity, as described in \citet{direwolf_paper}. Four uncertain parameters are considered: (1) $Q_{evap}$: the power removed by (or heating rate of) the HP; (2) $T_{sink}$: the sink temperature on the HP condenser; (3) ${htc}_{sink}$: the corresponding heat transfer coefficient; and (4) $R_{pore}$: the pore radius in the HP wick. Each of these parameters was assumed to follow normal distributions, with the means being defined consistently with what was used in \citet{Terlizzi2023} (i.e., 1821~W, 900~K, $10^3$ W/K/m$^2$, and 45 $\mu$m, respectively). The standard deviation for each parameter was arbitrarily chosen to be equal to 10\% of the mean.

We used three forward UQ approaches in MOOSE to quantify the low probability of HP failure: (1) Monte Carlo, which serves as the reference solution but is computationally expensive; (2) standard subset simulation executed in a massively parallel fashion; and (3) subset simulation with active learning via a standard GP. These approaches are detailed below.
\begin{itemize}
    \item \textit{Monte Carlo:} We used $10^9$ MOOSE model evaluations to compute the HP failure probability.
    \item \textit{Standard subset simulation executed in parallel:} We used seven subsets and $20,000$ MOOSE model evaluations per subset. In each subset, we used 40 independent Markov chains, each evaluating the MOOSE model 500 times in serial. These 40 Markov chains were launched in parallel fashion. Intermediate thresholds were computed, corresponding to a probability of 0.1. In total, the MOOSE model was evaluated $140,000$ times to compute the failure probability.
    \item \textit{Active learning subset simulation:} We used seven subsets and $2,000$ samples per subset. The input samples were first evaluated by using a standard GP to predict the MOOSE model output. If the GP prediction, as deemed by the U-function (see Table \ref{tab:MOOSE_acq_functions}), is inadequate, only then is the MOOSE model evaluation performed. Intermediate thresholds were computed, corresponding to a probability of 0.1. Note that the number of actual MOOSE model evaluations depends on the adequacy of the GP model for each input sample. This is discussed in detail next.
\end{itemize}

\begin{figure}[htb!]
    \centering  
    \includegraphics[scale=0.65]{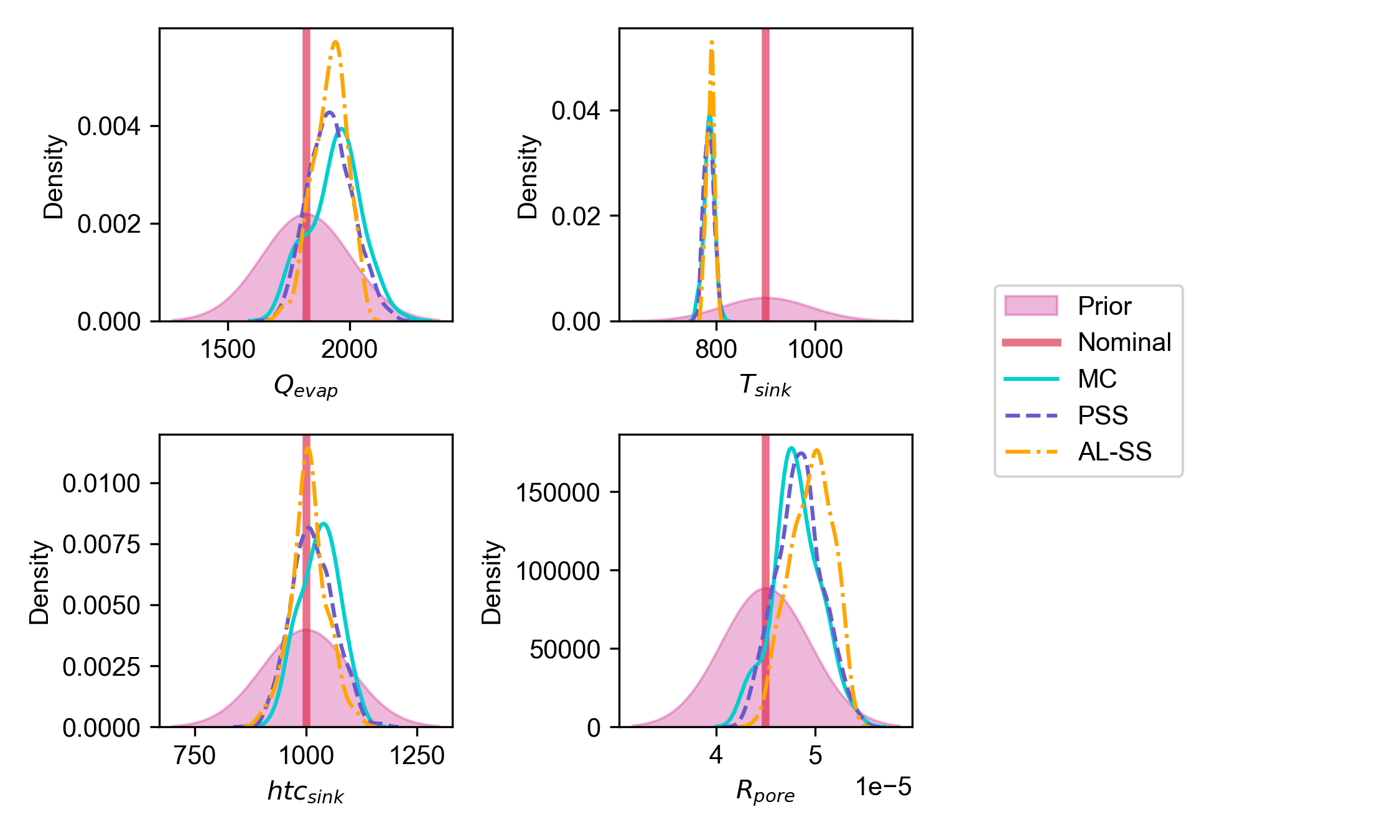}
    \caption{Distributions of the input parameters for failed HPs when comparing the three approaches: Monte Carlo (MC), parallelized subset simulation (PSS), and active learning subset simulation (AL-SS). Also shown for reference are the nominal input parameter distributions to the MOOSE HP model.}
    \label{fig:comparison_distribution_params_hp}
\end{figure}

Table \ref{tab:HP_results} presents the failure probabilities computed using the three different approaches, along with the corresponding coefficient of variation, the total number of MOOSE model evaluations, and the required number of processors. First, note that all three methods return similar failure probability values. As the failure probability is extremely small, Monte Carlo requires an enormous number of MOOSE model evaluations. Subset simulation reduces this number by a factor of $7,000$ as compared to Monte Carlo. Active learning subset simulation reduces this number even further, by a factor of $7.7 \times 10^6$ and $1,000$ in comparison to Monte Carlo and subset simulation, respectively. Figure \ref{fig:comparison_distribution_params_hp} presents the distributions of input parameters for failed HPs so as to enable further comparison of the three approaches. Note that all three return similar input parameter distributions for the failed HPs.

\begin{table}[htb!]
\centering
\caption{Comparison of the statistics for the three forward UQ approaches in MOOSE when evaluating the failure of an HP microreactor model. Shown for reference are the number of MOOSE model evaluations and the number of required processors utilized when computing the failure probabilities.}
\label{tab:HP_results}
\small
\begin{tabular}{ |c|c|c|c|c| }
\hline
\textbf{Method} &  \Centerstack{\textbf{Failure} \\ \textbf{probability}} & \Centerstack{\textbf{Coefficient of} \\ \textbf{variation}} & \Centerstack{\textbf{MOOSE model} \\ \textbf{evaluations}} & \textbf{Processors required}\\
\hline
\Centerstack{Monte \\ Carlo} & $7 \times 10^{-8}$ & $0.12$ & $10^9$ & 192\\
\hline
\Centerstack{Parallelized \\ subset simulation} & $5.1\times 10^{-8}$ & $0.06$  & $140,000$ & 40 \\
\hline
\Centerstack{Active learning \\ subset simulation} & $4.75\times 10^{-8}$ & $0.192$ & $130$ & 1 \\
\hline
\end{tabular}
\begin{tablenotes}
\small
\item[] The ``MOOSE model evaluations" column represents the total number of model evaluations required.
\end{tablenotes}
\end{table}
\normalsize

\subsection{Multi-output Gaussian processes and dimensionality reduction for advanced manufacturing simulations}\label{sec:AM}

This section uses the MOGP and dimensionality reduction capabilities discussed in Sections \ref{sec:MOGP} and \ref{sec:dimRed}, respectively.

\begin{figure}[htb!]\centering
          \includegraphics[width=0.7\textwidth]{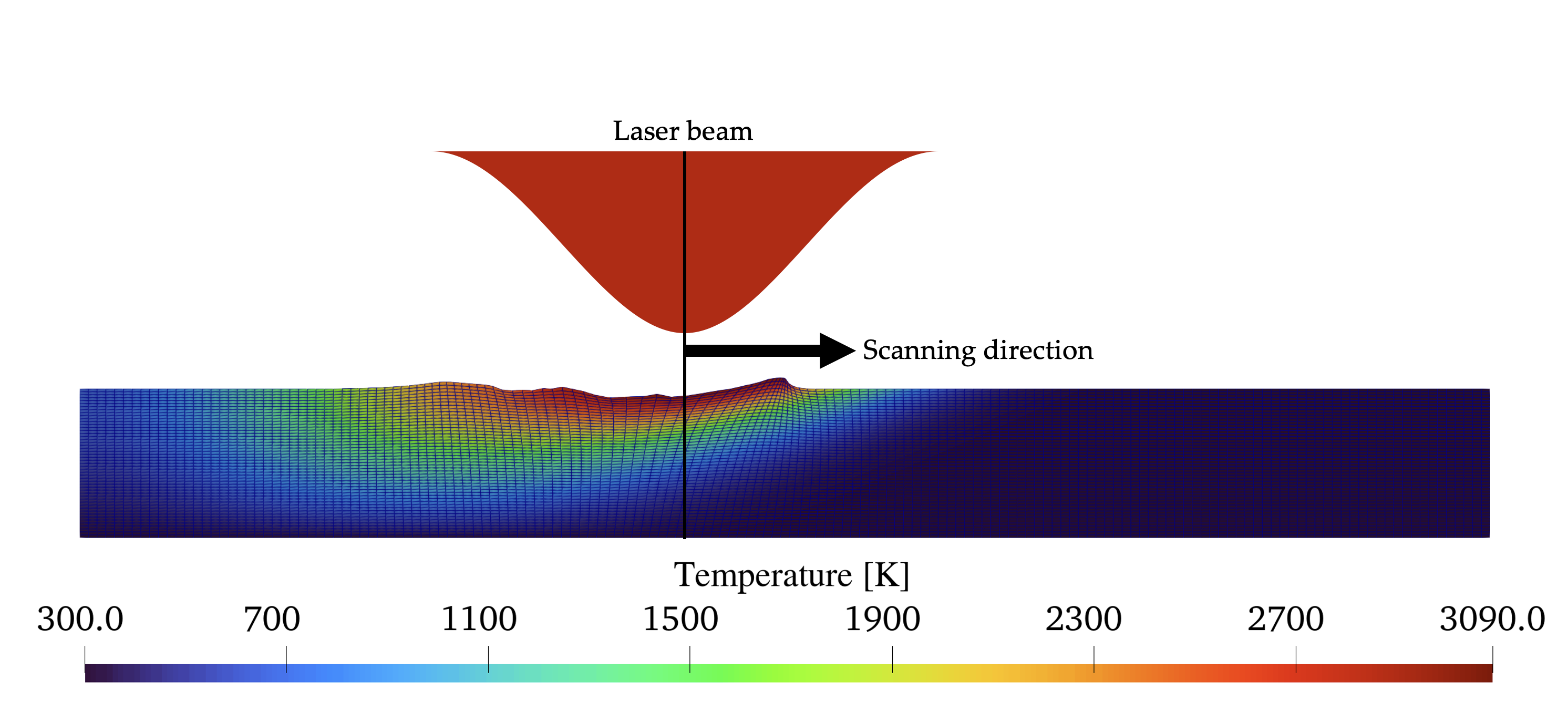}
 \caption{Temperature field output from the high-fidelity MOOSE model, which simulated the advanced manufacturing process by considering an effective laser power of $P=81.97~W$ and a laser radius of $R=125.8~\mu m$. The model relied on the Arbitrary Lagrangian-Eulerian method for capturing deformations caused by the vapor pressure on the melt pool surface.}
 \label{fig:test_T}
\end{figure}

Several advanced manufacturing techniques, including direct energy deposition and laser powder bed fusion, rely
on the melting of metals with the help of a laser. The quality of the final product depends on the process parameters employed (e.g. laser power and beam radius). However, simulation of laser melt pools is challenging due to the multiple physics involved, including melting and solidification along with fluid dynamics and heat transfer in the melt pool. This is why development of surrogate models for such simulations carries high potential for accelerating parametric studies that aim to explore the relationship between process parameters and product quality. We trained an MOGP-based surrogate model combined with dimensionality reduction, using linear PCA within MOOSE to predict full temperature fields during the advanced manufacturing process \cite{Biswas2024malamute}. The high-fidelity MOOSE model was run to gather temperature fields with different process parameters---namely, effective laser power and effective laser beam radius. The MOOSE model relied on the Arbitrary Lagrangian-Eulerian method for capturing deformations caused by the vapor pressure on the melt pool surface. Figure~\ref{fig:test_T} presents the MOOSE model setup, together with the temperature distribution for a specific combination of the two process parameters. 

\begin{figure}[htb!]
    \centering
    \begin{subfigure}[b]{0.49\textwidth} \centering
        \includegraphics[width=1\textwidth]{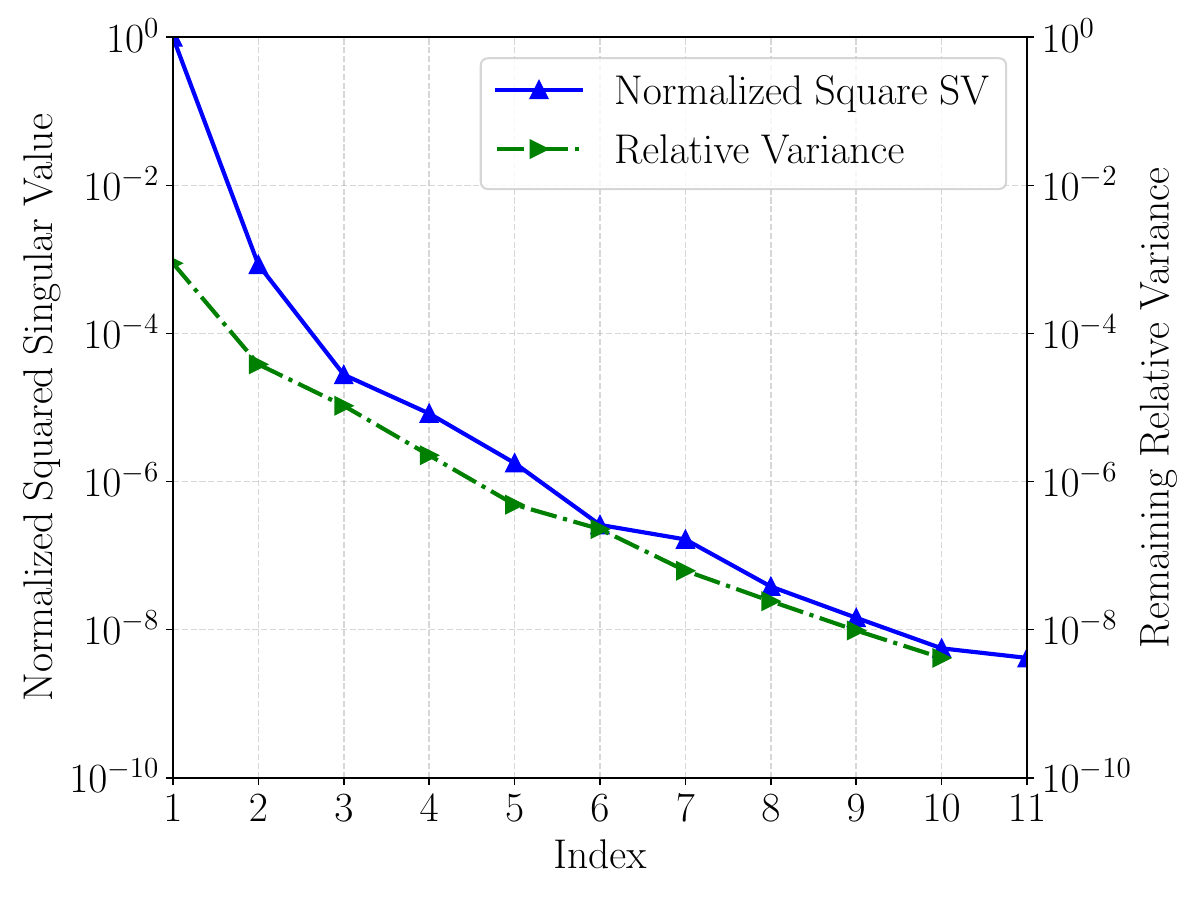}
    \caption{}
    \label{fig:scree_T}
    \end{subfigure}
    \hfill
    \begin{subfigure}[b]{0.49\textwidth} \centering
        \includegraphics[width=1\textwidth]{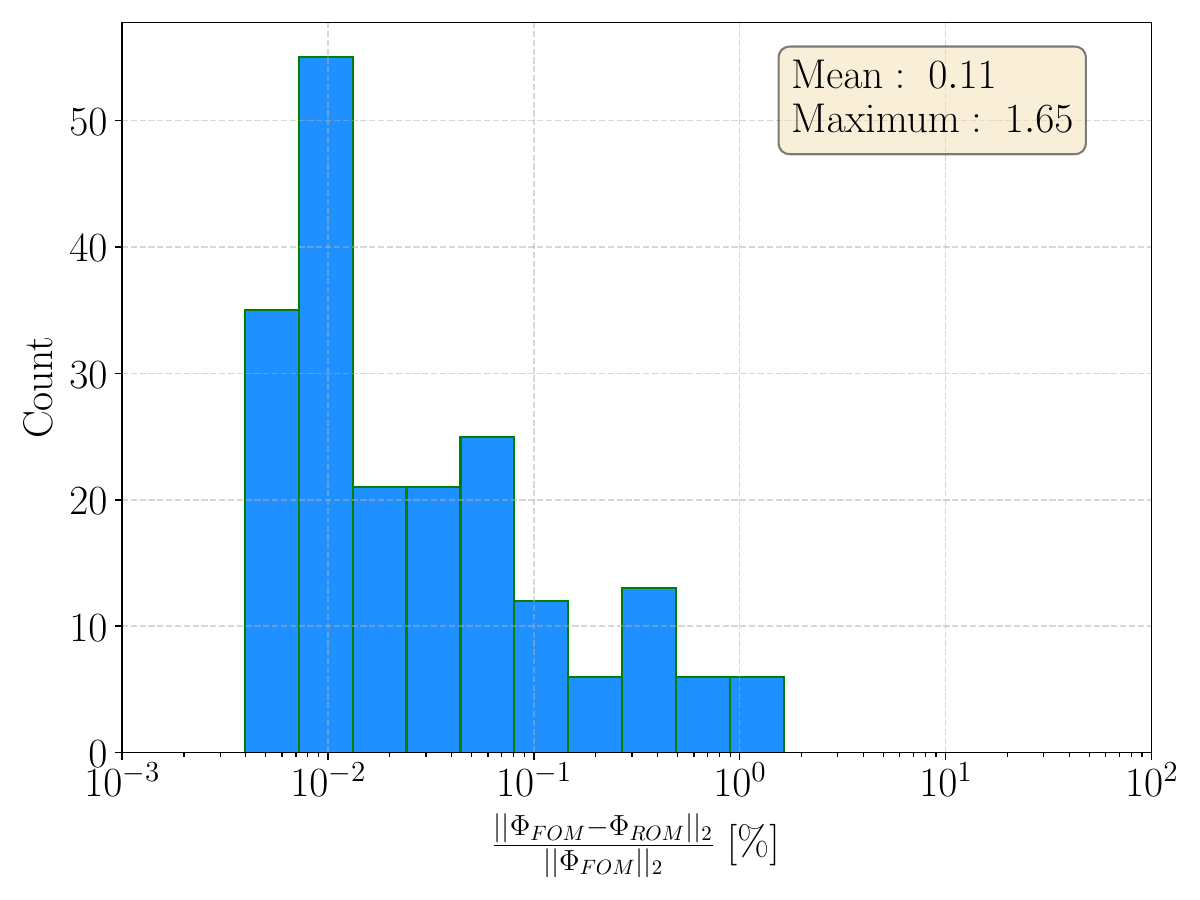}
        \caption{}
        \label{fig:errorhist_T}
    \end{subfigure}
    \label{fig:MOGP_error}
    \caption{(a) Decay in the squared singular values of the temperature fields upon performing linear PCA. The remaining relative variance is also shown, as computed by excluding the variance of the modes up the given index. (b) Histogram of the relative $L^2$ errors (in \%) of the temperature field between the high-fidelity model and the reconstructed solution from the MOGP by considering the testing set of 200 samples.}
\end{figure}

In this work, the temperature field at a given time step was the primary quantity of interest. In total, 120 snapshots of temperature fields were collected from the high-fidelity model by varying the process parameters. LHS was employed to randomize the process parameters, using $\mathcal{U}(70,83)$ [W] and 
$\mathcal{U}(125,200)~\left[\mu \textrm{m}\right]$ for the effective laser power and beam radius, respectively. Then linear PCA was applied to the temperature field snapshots for data compression. The decay of the squared singular values and the relative variance content are presented in Figure~\ref{fig:scree_T}. We see rapid decay in the explained variance, indicating that a few PCA components are sufficient to describe the thermal behavior of the system. Based on this information, a latent space of 10 dimensions was selected, and the temperature snapshot fields were mapped onto this space by using the first 10 components of linear PCA.

\begin{figure}[htb!]\centering
          \includegraphics[width=0.7\textwidth]{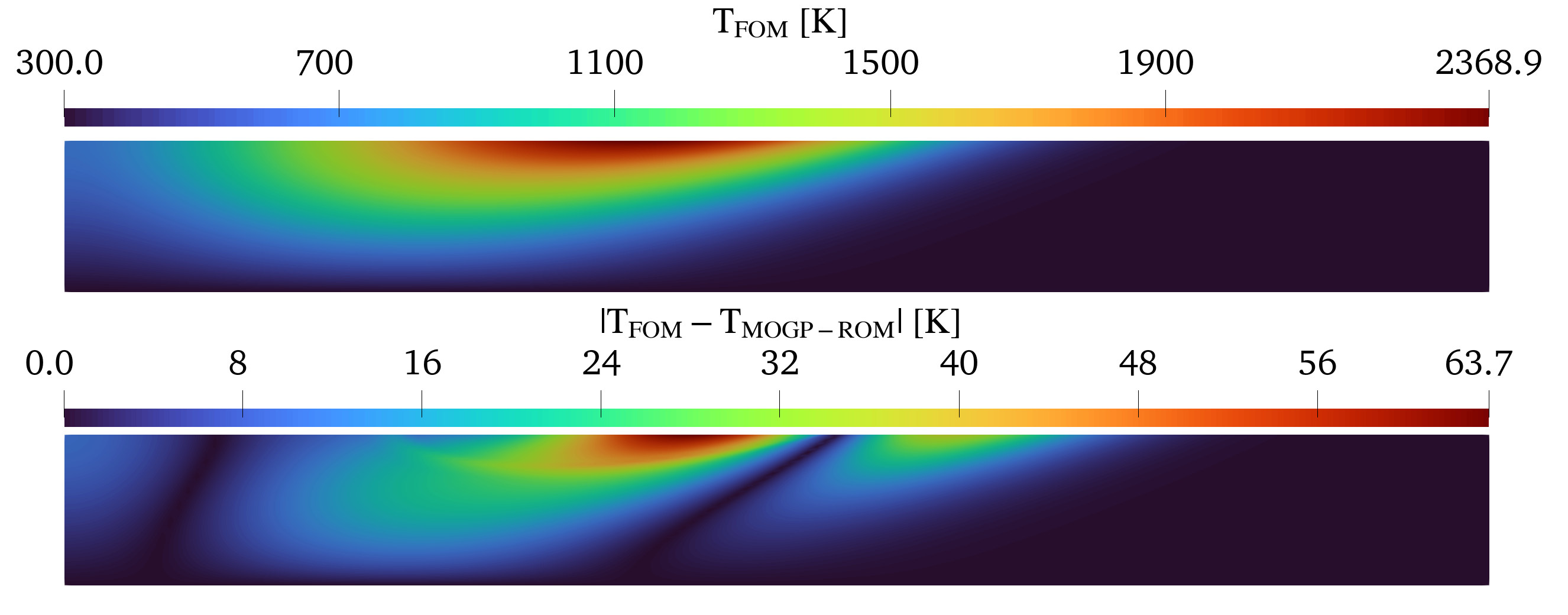}
 \caption{Comparison of solutions from the high-fidelity MOOSE model and reduced-order models at the least accurate sample in the testing set. Top: temperature profile computed using the high-fidelity MOOSE model. Bottom: absolute difference between the MOOSE model and the reconstructed MOGP solutions.}
 \label{fig:error_example_T}
\end{figure}

The 10 latent space components across 120 random realizations of the process parameters served as the training samples for the MOGP. The MOGP was trained via Adam optimization with 1,000 epochs, at a learning rate of $5\times10^{-4}$. The trained MOGP was then evaluated on a test set, using 200 samples of the process parameters. The MOGP-predicted latent quantities, which have 10 dimensions, were projected back to the original space by using an inverse PCA. The reconstructed temperature fields were then compared against the reference temperature fields obtained by evaluating the high-fidelity MOOSE model. Figure~\ref{fig:errorhist_T} presents a histogram of the relative full-field errors (in percentages) for the testing set. Generally, these relative errors are quite small, with a mean relative error of around 0.1\% and a maximum relative error of 1.65\%. The maximum error occurs near the boundary of the parameter domain, which was not properly covered by the training set, thus leading to minor inaccuracies in the MOGP prediction. The reference temperature field, along with the space-dependent absolute error between the reference and the MOGP solutions, is presented in 
Figure~\ref{fig:error_example_T} for those process parameters with the highest relative error in Figure \ref{fig:errorhist_T}. We see that the highest space-wise error is approximately 2.5\%, which is acceptable for the given use case. Evaluation of the MOGP occurred 4--5 orders of magnitude faster than the solving of the transient melt pool simulation. This further reflects the high potential for accelerating parameter studies related to the product quality's dependence on process parameters, in addition to permitting active learning based on the uncertainty estimates of the MOGP.

\subsection{Comparing deep and standard Gaussian processes for a lid-driven cavity flow}\label{sec:fluid}

This section uses the GP and deep GP capabilities discussed in Sections \ref{sec:GP} and \ref{sec:DGP}, respectively.

In this section, we compare a DGP trained using MCMC against a standard GP trained using either MCMC or Adam optimization for a four-sided lid-driven cavity flow problem. The fluid domain, a 2D square region defined by viscosity and density, is subjected to velocity boundary conditions on all four sides. The pressure is set to zero at the lower-left corner. More details on the problem setup can be found in \citet{Dhulipala2022c}. We are interested in predicting the resultant velocity at the domain's center as a function of the viscosity and density of the fluid and of the four boundary conditions. We used the MOOSE Navier-Stokes Module to generate training and testing data under random values for the viscosity and density of the fluid and the four boundary conditions \cite{Peterson2018a}. 

The training data were comprised of 30 points, and the testing data were comprised of 100. We first trained a standard GP by using MCMC. There were seven hyperparameters to optimize (i.e., six length scales and one amplitude scale), and we used 10,000 samples in the MCMC algorithm in order to estimate the posterior distributions of the hyperparameters. We then trained a DGP with one hidden layer using MCMC. This time, there were 43 hyperparameters; that is, six for each of six nodes in the hidden layer, plus an additional seven for the output layer. We again used 10,000 samples in the MCMC algorithm so as to estimate the posterior distributions of the hyperparameters. Finally, we trained a standard GP by using Adam optimization (giving us seven hyperparameters to optimize). The Adam optimization entailed 1,000 iterations, a learning rate of 0.005, and a batch size of 20. 

We compared the three approaches for predicting the resultant velocity---namely, GP using MCMC, DGP using MCMC, and GP using Adam optimization---based on diagnostics such as parity plots, calibration curves, uncertainty distributions, and error bars, as detailed in \citet{Tran2020a,Kuleshov2018a}. The parity plots assessed the accuracy of the predictions and presented metrics such as median absolute error, root mean squared error, mean absolute error, and mean absolute relative percent difference. Calibration curves ``use the standard deviation predictions to create Gaussian random variables for each test point and then test how well the residuals followed their respective Gaussian
random variables'' \cite{Tran2020a}. In other words, the model is said to be well calibrated if the expected-vs.-observed cumulative distribution of the testing points follows a straight line. A well-calibrated model could still have large uncertainty estimates that are less useful in practice \cite{Tran2020a}. Thus, from the uncertainty distributions, metrics such as sharpness and coefficient of variation $(C_v)$ are derived. Large uncertainty estimates are less desirable than small values, and sharpness assesses this by taking the root mean of the predicted variances. The model should not predict constant uncertainty estimates outside the training bounds, and $C_v$ assesses this by computing the coefficient of variation of the predictive variances. While smaller values of the accuracy metrics, miscalibration area, and sharpness are preferred, a larger value of $C_v$ is desirable.    

\begin{figure}[htb!]
\centering
    \begin{subfigure}[b]{0.24\textwidth} \centering
        \includegraphics[width=1.1\linewidth]{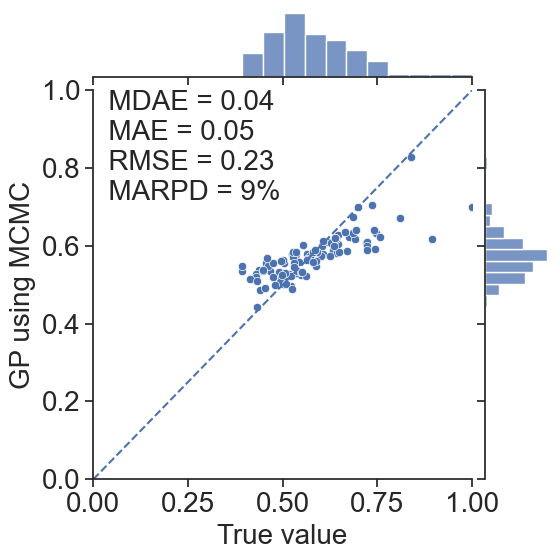}
    \caption{}
    \label{fig:gpmc_parity}
    \end{subfigure}
    \hfill
    \begin{subfigure}[b]{0.24\textwidth} \centering
        \includegraphics[width=1\linewidth]{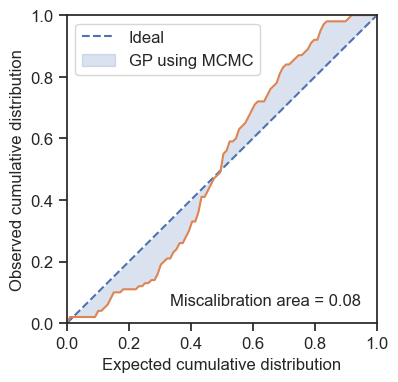}
    \caption{}
    \label{fig:gpmc_calibration}
    \end{subfigure}
    \hfill
    \begin{subfigure}[b]{0.24\textwidth} \centering
        \includegraphics[width=0.9\linewidth]{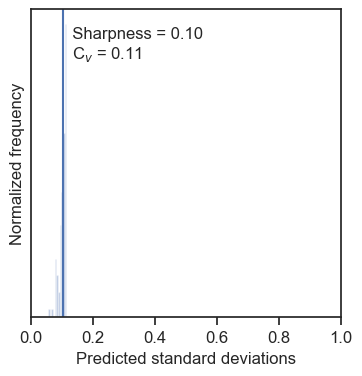}
    \caption{}
    \label{fig:gpmc_dispersion}
    \end{subfigure}
    \hfill
    \begin{subfigure}[b]{0.24\textwidth} \centering
        \includegraphics[width=1\linewidth]{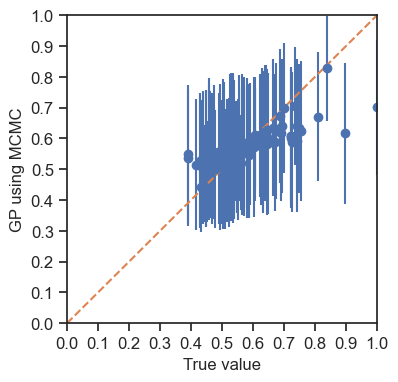}
    \caption{}
    \label{fig:gpmc_errorbar}
    \end{subfigure}
    \hfill
    \begin{subfigure}[b]{0.24\textwidth} \centering
        \includegraphics[width=1.1\linewidth]{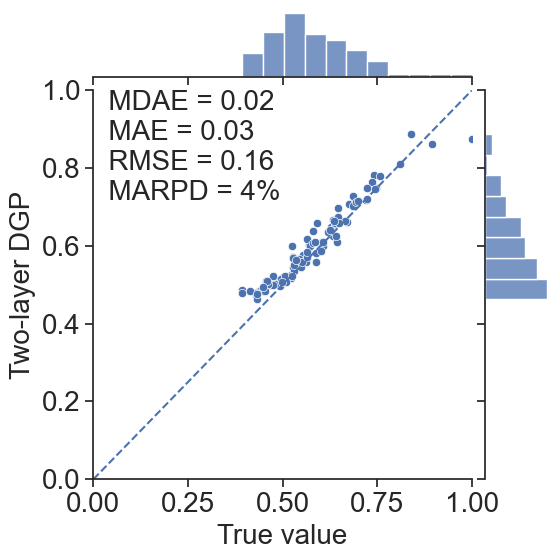}
    \caption{}
    \label{fig:tgp_parity}
    \end{subfigure}
    \hfill
    \centering
    \begin{subfigure}[b]{0.24\textwidth} \centering
        \includegraphics[width=1\linewidth]{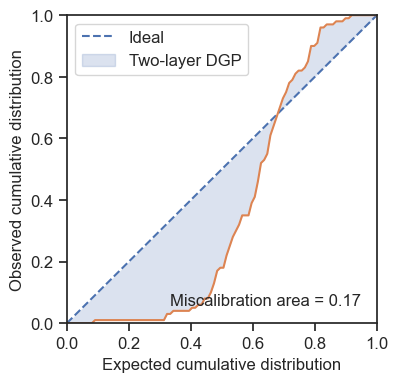}
    \caption{}
    \label{fig:tgp_calibration}
    \end{subfigure}
    \hfill
    \begin{subfigure}[b]{0.24\textwidth} \centering
        \includegraphics[width=0.9\linewidth]{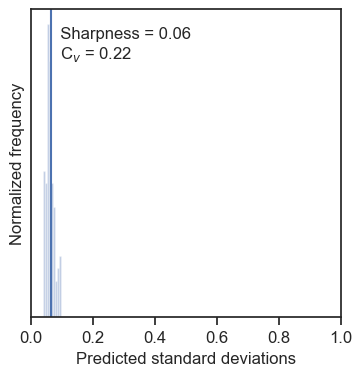}
    \caption{}
    \label{fig:tgp_dispersion}
    \end{subfigure}
    \hfill
    \begin{subfigure}[b]{0.24\textwidth} \centering
        \includegraphics[width=1\linewidth]{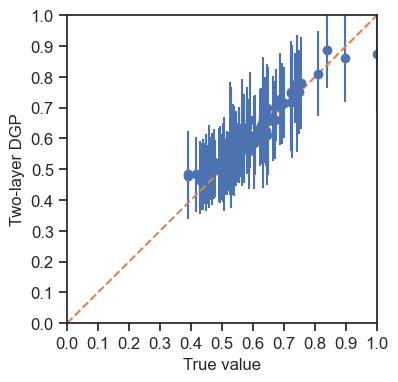}
    \caption{}
    \label{fig:tgp_errorbar}
    \end{subfigure}
    \hfill
    \begin{subfigure}[b]{0.24\textwidth} \centering
        \includegraphics[width=1.1\linewidth]{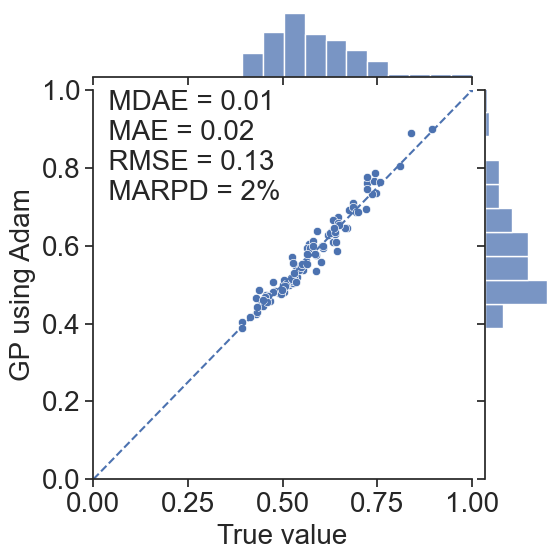}
    \caption{}
    \label{fig:gpadam_parity}
    \end{subfigure}
    \hfill
    \begin{subfigure}[b]{0.24\textwidth} \centering
        \includegraphics[width=1\linewidth]{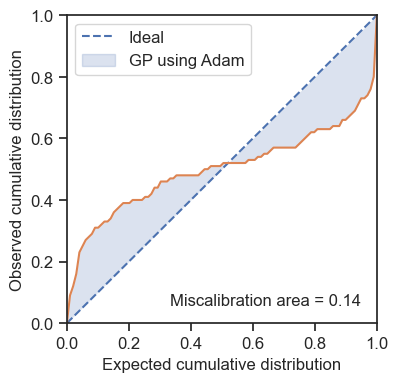}
    \caption{}
    \label{fig:gpadam_calibration}
    \end{subfigure}
    \hfill
    \begin{subfigure}[b]{0.24\textwidth} \centering
        \includegraphics[width=0.9\linewidth]{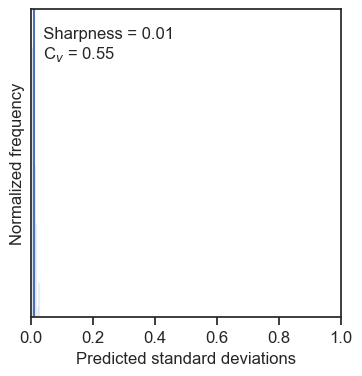}
    \caption{}
    \label{fig:gpadam_dispersion}
    \end{subfigure}
    \hfill
    \begin{subfigure}[b]{0.24\textwidth} \centering
        \includegraphics[width=1\linewidth]{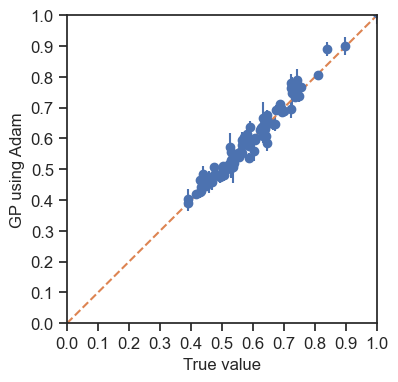}
    \caption{}
    \label{fig:gpadam_errorbar}
    \end{subfigure}
\caption{Predictive performance of the three GP variants in terms of both accuracy and uncertainty quality. Top row [(a)--(d)]: GP trained using MCMC. Middle row [(e)--(h)]: DGP trained using MCMC. Bottom row [(i)--(l)]: GP trained using Adam optimization. (a), (e), and (i) show the parity plots and present accuracy metrics such as median absolute error, root mean squared error, mean absolute error, and mean absolute relative percent difference. (b), (f), and (j) show the calibration plots and miscalibration area metric for uncertainty quality. (c), (g), and (k) show histograms of the predictive standard deviations, the metric sharpness, and the $C_v$. (d), (h), and (l) show the error bars.}
\label{fig:dgp_gp_1}
\end{figure}

Figure \ref{fig:dgp_gp_1} compares the three approaches---GP using MCMC, DGP using MCMC, and GP using Adam optimization---in light of the aforementioned diagnostics. The first row corresponds to GP using MCMC, the second row to DGP using MCMC, and the third row to GP using Adam optimization. In comparing GP using MCMC against DGP using MCMC, the latter generally outperforms the former in almost every metric. DGP using MCMC has better accuracy, lower sharpness, and a larger $C_v$ than GP using MCMC, showing the power of DGP method compared to GP. Although GP using MCMC has a smaller miscalibration area, this is likely due to it predicting constant wider uncertainty bands (as observed by comparing Figure \ref{fig:gpmc_errorbar} to Figure \ref{fig:tgp_errorbar}) than does DGP using MCMC. As such, hidden layers help a DGP model with more expressivity and better uncertainty quality than a standard GP when trained using MCMC. In comparing DGP using MCMC against GP using Adam optimization, the latter outperforms the former in every metric. We suspect that this is largely due to the inefficiency of MCMC in high-dimensional parameter spaces in DGP and the optimization algorithm plays a big role in the predictive performance (including accuracy and uncertainty quality) of the GP models. In the future, DGP will be implemented with a more efficient variational inference and gradient-based solvers coupled with MOOSE’s libtorch capabilities \cite{German2023s}.

\subsection{Batch Bayesian optimization of tritium diffusion experiment in beryllium}\label{sec:fusion}

This section uses the active learning with GP capabilities discussed in Section \ref{sec:AL}.

\begin{figure}[htb!]
    \centering
    \begin{subfigure}[b]{0.49\textwidth} \centering
        \includegraphics[width=1\textwidth]{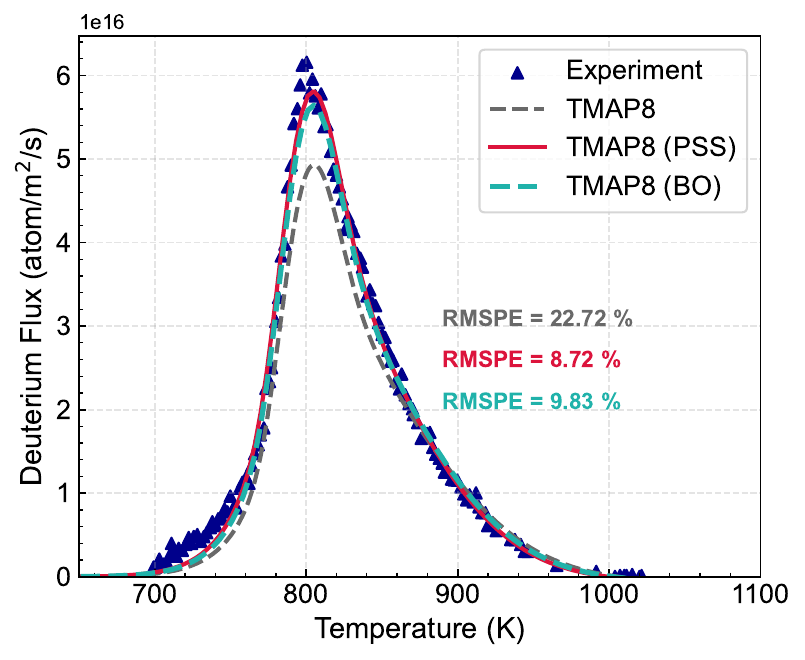}
    \caption{}
    \label{fig:tmap8_calibration}
    \end{subfigure}
    \hfill
    \begin{subfigure}[b]{0.49\textwidth} \centering
        \includegraphics[width=0.95\textwidth]{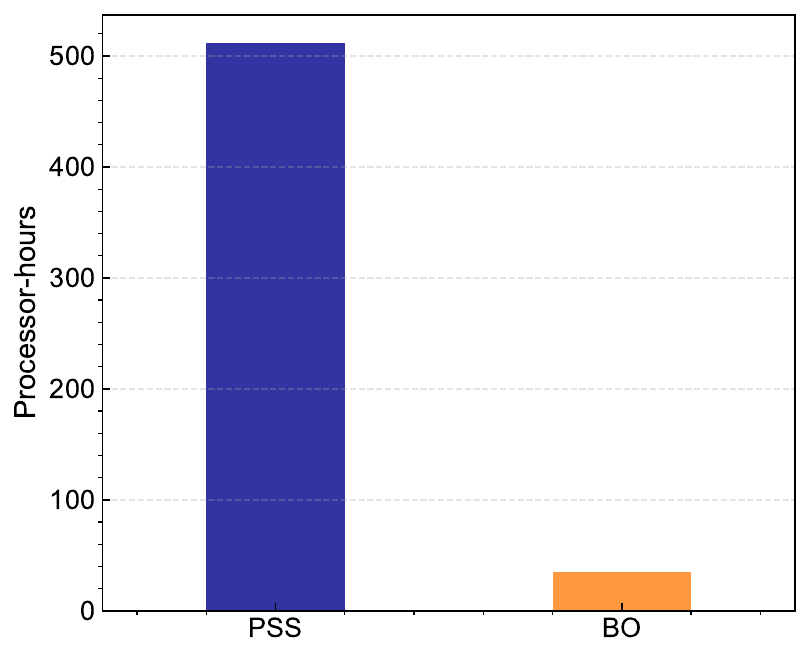}
        \caption{}
        \label{fig:tmap8_time}
    \end{subfigure}
    \label{fig:tmap8}
    \caption{(a) Modeled deuterium flux during desorption, compared against experimental data. Before calibrating the model parameters, the model had a root mean squared percent error of $22.72\%$ when examined against the experimental data. Upon calibration based on parallel subset simulation (an evolutionary approach) and batch Bayesian optimization, the root mean squared percent error reduced to $8.72\%$ and $9.83\%$, respectively. (b) The computational cost of calibrating the model parameters via parallel subset simulation and Bayesian optimization was measured as the product of the number of processors and the elapsed time (in hours).}
\end{figure}

The Tritium Migration Analysis Program, Version 8 (TMAP8) is a state-of-the-art, open-source, MOOSE-based application designed for multiscale tritium transport. TMAP8 incorporates multispecies, multiphysics, multiscale simulation capabilities on complex geometries. These capabilities make it an essential tool for the fusion energy community, particularly for addressing the challenges of tritium tracking, fusion system safety, and fuel sustainability. Validation case study val-2b in TMAP8's test suite validates against implantation and thermal absorption/desorption experiments on wafers of polished beryllium from \citet{Macaulay1991a}. The beryllium was exposed to 13.3 kPa of deuterium at 773 K for 50 hours, cooled down to 300 K in vacuum, and then heated back up to 1073 K at a rate of 3 K/min to desorb the deuterium. Further details are available in \citet{Simon2025}. The modeled deuterium flux during desorption was compared against experimental data, as shown in Figure \ref{fig:tmap8_calibration}. Herein, batch Bayesian optimization was applied to calibrate the diffusivities and solubilities of the TMAP8 model in order to improve agreement with the experimental data. 

The deuterium flux model had 10 parameters, comprised of the diffusivities and solubilities that must be calibrated against the experimental data. Prior to the calibration, the model predictions and the experimental data resulted in a root mean squared percent error of $22.72\%$. Two approaches in MOOSE were used to achieve this calibration: parallel subset simulation, which is an evolutionary approach, and batch Bayesian optimization. The aim of the optimization with respect to the model parameters was to minimize the mean squared percentage error between the experimental data and the model predictions regarding the deuterium flux during desorption. Details on the usage of these approaches are as follows:
\begin{itemize}
    \item \textbf{Parallel subset simulation:} Run for five subsets, with 1,000 samples per subset. Ten processors were used, simultaneously simulating five parallel chains for 1,000 serial model evaluations.
    \item \textbf{Batch Bayesian optimization:} Run for 80 serial iterations, with a batch of five optimal points selected in parallel in each iteration by using the expected improvement acquisition function. A standard GP with squared exponent covariance matrix was trained using Adam optimization, in which 2,000 iterations were performed using a learning rate of 0.01. The five selected optimal points were used for evaluating the computational model in parallel.
\end{itemize}
Figure \ref{fig:tmap8_calibration} presents the model output against the experimental data following the parameter calibration. We see that both the parallel subset simulation and batch Bayesian optimization have similar root mean squared percent error values, and both substantially reduce this error metric in comparison to the uncalibrated model. Figure \ref{fig:tmap8_time} presents the computational burden of the two approaches, as assessed based on the product of the number of processors and the elapsed time in hours. Ultimately, batch Bayesian optimization is revealed to be substantially lower in computational cost than parallel subset simulation.

\section{Summary and Conclusions}\label{sec:conc}

MOOSE, an open-source computational platform for parallel numerical analysis, is being actively developed and is maintained at Idaho National Laboratory. MOOSe has an extensive user base in varied scientific and engineering fields. Complex multiphysics simulations, when validated against experimental data, are subject to different sources of uncertainties that must be quantified and propagated to the outputs. They are also computationally expensive to run, especially in a UQ setting, and surrogate models for quantifying their prediction uncertainties will foster their efficient and accurate execution by leveraging active learning principles. In this context, the present paper covered the development and demonstration of massive parallel probabilistic ML and UQ capabilities in MOOSE. Among these capabilities are active learning, Bayesian inverse UQ, adaptive forward UQ, Bayesian optimization, evolutionary optimization, and MCMC. The MOOSE systems \texttt{Sampler}, \texttt{MultiApp}, \texttt{Reporter}, and \texttt{Surrogate}, as well as the modularity thereof, were discussed in detail in regard to successfully developing a multitude of probabilistic ML and UQ algorithms. Example code demonstrations include parallel active learning and parallel Bayesian inference via active learning. Finally, the impacts of these code developments were discussed in regard to five different applications: nuclear fuel fission product release, using parallel active learning Bayesian inference; nuclear microreactor very rare events analysis, using active learning; advanced manufacturing process modeling, using MOGP and dimensionality reduction; lid-driven cavity flow, using DGPs; and tritium transport for fusion energy, using batch Bayesian optimization.

\section*{Acknowledgements}
The forward UQ capability developments, including active learning and multifidelity modeling for forward problems, are supported through Idaho National Laboratory (INL)'s Laboratory Directed Research \& Development (LDRD) Program under U.S. Department of Energy (DOE) Idaho Operations Office Contract DE-AC07-05ID14517.

The Bayesian inverse UQ capability developments, including active learning for inverse problems, are supported through Battelle Energy Alliance, LLC under contract no. DE-AC07-05ID14517 with DOE, along with funding from the Nuclear Energy Advanced Modeling and Simulation (NEAMS) program within the DOE Office of Nuclear Energy (DOE-NE).

The multi-output Gaussian processes and dimensionality reduction capability developments are supported through Battelle Energy Alliance, LLC under contract no. DE-AC07-05ID14517 with DOE, with funding from the Advanced Materials and Manufacturing Technologies (AMMT) program within DOE-NE.

The deep Gaussian processes and Bayesian optimization capability developments are supported through Battelle Energy Alliance, LLC under contract no. DE-AC07-05ID14517 with DOE, along with funding from the Nuclear Energy University Partnerships (NEUP) program within DOE-NE.

This research made use of resources of the High-Performance Computing Center at INL, which is supported by DOE-NE and the Nuclear Science User Facilities under contract no. DE-AC07-05ID14517.

We thank the following individuals for their support in developing the capabilities of the MOOSE Stochastic Tools Module: Stephen R. Novascone, Sudipta Biswas, Benjamin W. Spencer, Jason D. Hales, and Daniel Schwen from Idaho National Laboratory; Michael D. Shields and Promit Chakroborty from Johns Hopkins University; and Andi Wang from the University of Wisconsin-Madison. We thank John Shaver at INL for his technical edit of this paper.

\bibliography{biblio}

\end{document}